%% file: postprocessingArXivFinal.tex
\documentclass[sigconf]{acmart}
\usepackage{bbm}
\AtBeginDocument{%
  }

\copyrightyear{2023} 
\acmYear{2023} 
\setcopyright{acmlicensed}\acmConference[CCS '23]{Proceedings of the 2023 ACM SIGSAC Conference on Computer and Communications Security}{November 26--30, 2023}{Copenhagen, Denmark}
\acmBooktitle{Proceedings of the 2023 ACM SIGSAC Conference on Computer and Communications Security (CCS '23), November 26--30, 2023, Copenhagen, Denmark}
\acmPrice{15.00}
\acmDOI{10.1145/3576915.3623081}
\acmISBN{979-8-4007-0050-7/23/11}

\usepackage{balance}
\usepackage{xcolor}
\usepackage{amsmath, amsthm, amsfonts}
\usepackage{stmaryrd} 
\usepackage{algorithm}
\usepackage{algorithmicx}
\usepackage{algpseudocode}
\usepackage{comment}
\usepackage{wrapfig}

\definecolor{ForestGreen}{rgb}{0.13, 0.54, 0.13}

\newcommand\CStuff[1] {\textcolor{blue}{#1}}
\newcommand\Cz {\CStuff{$\star$\,}}

\newcommand\Cf[1] {\CStuff{$|$}\marginpar{\CStuff{\fbox{\footnotemark}}}\footnotetext{\CStuff{#1}}}
\newcommand\AStuff[1] {\textcolor{red}{#1}}
\newcommand\Az {\AStuff{$\star$\,}}
\newcommand\AMA {\ensuremath{\mathit{ama}}}
\newcommand\Af[1] {\AStuff{$|$}\marginpar{\AStuff{\fbox{\footnotemark}}}\footnotetext{\AStuff{#1}}}
\newcommand\NStuff[1] {\textcolor{orange}{#1}}

\newcommand\Nz {\NStuff{$\star$\,}}
\newcommand\Nf[1] {\NStuff{$|$}\marginpar{\NStuff{\fbox{\footnotemark}}}\footnotetext{\NStuff{#1}}}
\newcommand\GStuff[1] {\textcolor{teal}{#1}}
\newcommand\Gz {\GStuff{$\star$\,}}
\newcommand\Gf[1] {\GStuff{$|$}\marginpar{\GStuff{\fbox{\footnotemark}}}\footnotetext{\GStuff{#1}}}
\newcommand\MStuff[1] {\textcolor{magenta}{#1}}

\newcommand\Mf[1] {\MStuff{$|$}\marginpar{\MStuff{\fbox{\footnotemark}}}\footnotetext{\MStuff{#1}}}

\renewcommand\MStuff[1] {{#1}}
\renewcommand\CStuff[1] {{#1}}
\renewcommand\NStuff[1] {{#1}}
\renewcommand\AStuff[1] {{#1}}

\usepackage[nice]{nicefrac}
\usepackage{array}
\usepackage{float}
\usepackage{subcaption}

\newcommand\AandCStart {\bgroup\Grey}
\newcommand\AandCEnd {\egroup}

\renewcommand\Cf[1] {\relax}
\renewcommand\Af[1] {\relax}
\renewcommand\Mf[1] {\relax}
\renewcommand\Nf[1] {\relax}
\renewcommand\Gf[1] {\relax}

\input{macros}

\renewcommand\Label[1] {\label{#1}}
\begin{document}

\title[A Novel Analysis of Utility in Privacy Pipelines]{A Novel Analysis of Utility in Privacy Pipelines, \\ Using Kronecker Products and Quantitative Information Flow}



\author{M\'{a}rio S.\ Alvim}
\affiliation{%
	\institution{UFMG}
	\city{Belo Horizonte}
	\country{Brazil}
}
\email{msalvim@dcc.ufmg.br}

\author{Natasha Fernandes}
\affiliation{%
  \institution{Macquarie University}
  \city{Sydney}
  \country{Australia}
}
\email{natasha.fernandes@mq.edu.au}

\author{Annabelle McIver}
\affiliation{%
	\institution{Macquarie University}
	\city{Sydney}
    \country{Australia}
}
\email{annabelle.mciver@mq.edu.au}

\author{Carroll Morgan}
\affiliation{%
 \institution{UNSW and Trustworthy Systems}
  \city{Sydney}
\country{Australia}
}
\email{carroll.morgan@unsw.edu.au}

\author{Gabriel H.\ Nunes}
\affiliation{%
  \institution{Macquarie University and UFMG}
  \city{Sydney and Belo Horizonte}
  \country{Australia and Brazil}
}
\email{ghn@nunesgh.com}


\begin{abstract}
We combine Kronecker products, and quantitative information flow, to give a novel formal analysis for the fine-grained verification of \emph{utility} in complex privacy pipelines. The combination explains a surprising anomaly in the behaviour of utility of privacy-preserving pipelines --- that sometimes a \emph{reduction} in privacy results also in a \emph{decrease} in utility. We use the standard measure of utility for Bayesian analysis, introduced by Ghosh at al.\ \cite{Ghosh2009}, to produce tractable and rigorous proofs of the fine-grained statistical behaviour leading to the anomaly. More generally, we offer the prospect of formal-analysis tools for utility that complement extant formal analyses of privacy. We demonstrate our results on a number of common privacy-preserving designs.
\end{abstract}


\ccsdesc[??]{Formal Methods}
\ccsdesc[???]{Security and Privacy}


\maketitle

{ 
}

\section{Introduction}\Label{s1243X}

\Mf{I'm collecting here some things we should pay attention to in the final version of the paper.
\begin{enumerate}
	\item For names/mnemonics, now we are using:
	\begin{enumerate}
		\item $C$ for the perturbation/sanitization mechanism (there is no mnemonic here).
		\item $P$ for the post-processing step (which is hard to remember, since $P$ could stand for ``perturbation'', ``post-processing'' or ``pipeline'' equally).
		\item $M$ for the overall pipeline (although $M$ may sound like a mnemonic for ``mechanism'', which we use only for the perturbation/sanitization step).
	\end{enumerate}
	\MStuff{
	I think that the choices above are
	a little confusing.
	Perhaps better names/mnemonics that avoid such clashes would be:
	\begin{enumerate}
		\item $S$ for sanitisation mechanism (``sanitisation'' is more general and accurate than ``perturbation'', and it avoids the clash with all other ``p''-words)?
		\item $P$ for post-processing?
		\item $D$ for data-release workflow ($S$anitization followed by $P$ost-processing, and voiding the use of ``pipeline'' altogether)?
	\end{enumerate}
	I think that would help in the readability of the paper, but we have to check for further conflicts with other notation later in the paper.
	In case we agree on trying new names, I volunteer to make the changes.}
	\MStuff{
	\item The use of ``mechanism'' instead of ``distribution'' when appropriate (as pointed out by Natasha).
	\item The use of ``data consumer'' vs. ``data analyst''. We should probably stick to only one.
	\item Make sure every definition (and result) has a name, and that the use of emphasis on the defined term is uniform.
	\item What are the genders of: adversary, data subject, data analyst/consumer?
	We are using ``she'', ``he'' and ``they'' inconsistently.
	}
\end{enumerate}}

In the release of statistical information there are
two crucial, and often opposing,
goals:
on the one hand, ``privacy'' concerns the protection of sensitive data from those who should \emph{not} access them; on the other hand ``utility'' concerns the preservation of the value of those data to those who \emph{should} have access to them.
Because those goals are so frequently conflicting,
much effort has been dedicated to understanding and optimising
the trade-off between them~\cite{ASIKIS2020488,Li2009OnTT,pmlr-v108-geng20a,DBLP:conf/concur/AlvimFMN20}.

Many recent mathematical formulations of privacy
preservation for data-release mechanisms
employ some variant of differential-privacy~\cite{Dwork2013,Ding_2018,Rappor}.
The utility of such mechanisms is then formalised
as the expected loss (or gain) of a Bayesian decision-maker who combining the observation of
the mechanism's outputs with prior knowledge, strives
to extract information optimally~\cite{Ghosh2009,Nissim:10:FOCS}.
In such scenarios, privacy practitioners commonly
rely (even if implicitly) on the assumption
that the parameter $\VE$ of a differentially-private
mechanism controls the trade-off between privacy
and utility: in particular,
the higher the value of $\VE$, the more the output of the mechanism resembles the original data, and consequently the less private and more useful a data release is expected
to be \cite{Dwork2013}.  In practice, working under that assumption, one expects a monotone relationship between differential privacy's $\epsilon$ and utility, so that it can be ``tuned'' to suit the goals of an acceptable level of privacy versus a desired level of utility (or accuracy). 


Here we call that assumption \emph{stability}; and --surprisingly-- there are many common situations where \textbf{stability does not hold}. 
In particular we have found that there are cases of \emph{instability}, in which a decrease in the value of
$\VE$ can improve both privacy and utility
\emph{at the same time}.
The phenomenon was first identified ``in the wild''
from our reproduction of a typical application of differential privacy to real-world data (\Sec{ss1554}): we tried therefore to explain it.

Indeed our major goal was to isolate and characterise the conditions that distinguish stable from unstable scenarios; and to do that
we propose a novel formal analysis method based on Kronecker products and quantitative information flow. We found a scalable reasoning technique that, facilitating fine-grained models of privacy and utility, made it possible to understand the trade-off between the two.

\begin{figure}[tb]
	\flushleft 
	\includegraphics[width=\linewidth]{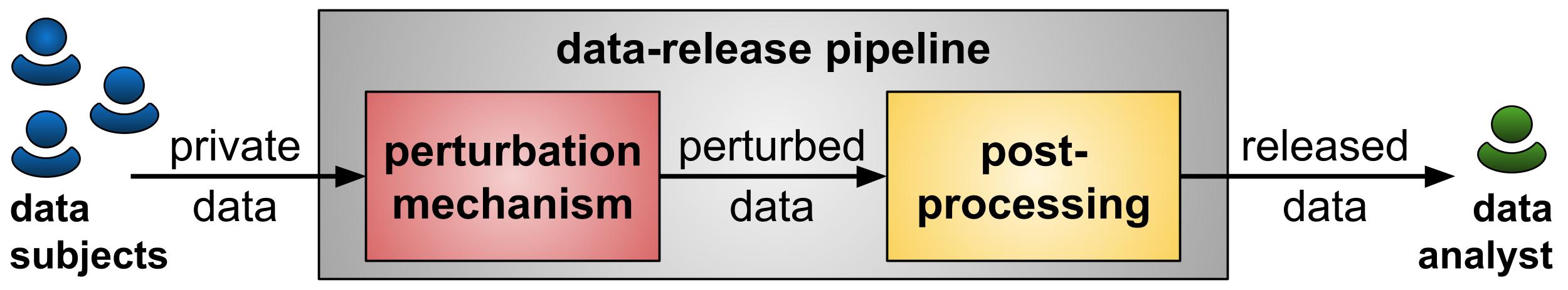}
This end-to-end depiction of a ``pipeline'' captures a typical architecture in which individuals' data is collected in sanitised form by first being obfuscated using a perturbing mechanism; the perturbed data is then subjected to analysis via some post-processing routine, such as count or sum querying.	 Whilst privacy is guaranteed for the individuals, the analyst would like to understand the impact on the accuracy of querying.
	\caption{Data-release pipeline considered in this work.
	}
	\label{fig:post-processing-pipelineX}
\end{figure}

We focus on
``data-release pipelines'' structured
as in \Fig{fig:post-processing-pipelineX}: they
capture many relevant real-world scenarios.
First the original, private data are collected from data subjects and sanitised by a perturbation mechanism.
Then a
post-processing step is applied to the perturbed data,
before release to the data analyst.
For example, in a statistical-database setup, the perturbation mechanism might correspond to a ``local'' differential privacy mechanism, i.e.\ one applied by each individual to their own data before sending it to a curator; and the subsequent post-processing is a statistical query performed by the analyst against the database \cite{doi:10.1137/090756090}. In a federated-learning setup however, the perturbation mechanism is applied by each client to the gradients of the model before submitting them to a central server; and the post-processing is an averaging applied by the server to the noisy gradients, which is then reported back to the clients \cite{yousefpour2022opacus}.   We remark that although these examples capture the so-called ``local model'' in differential privacy, our analysis is more abstract: it relies only on a perturbation followed by a post-processing step, and therefore could be applied in a variety of scenarios not usually associated with local differential privacy.

\Nf{Perhaps use aggregation in place of statistic. \MStuff{Hasn't this been addressed now that we changed it to ``data-release pipeline''?}}

Our formal model for such pipelines is based on
\textit{quantitative information flow} (\QIF),
which employs rigorous information-- and decision-theoretic principles
to reason about the amount of information that flows from inputs to outputs
in a computational system.
\QIF\ has been successfully applied already to a variety of privacy and security analyses,
including searchable encryption~\cite{jurado2021formal},
intersection- and linkage attacks against $k$-anonymity~\cite{fernandes2018processing},
privacy analysis of very large datasets~\cite{Alvim:22:PoPETSa},
and differential privacy~\cite{Alvim:15:JCS,Chatzi:2019}.
Although \QIF\ is usually employed to measure the flow of \emph{sensitive}
information through a system, i.e.\ the system's \emph{leakage},
in this work we employ the framework to assess the flow of \emph{useful}
information through a data-release pipeline,
i.e.\ its decision-theoretic \emph{utility} to a Bayesian data-analyst.
\MStuff{That notion of utility follows standard principles from decision theory~\cite{Degroot:1970:Book}, and
has been shown~\cite{DBLP:conf/csfw/FernandesMPD22} to be equivalent to influential
measures of utility in the privacy literature, such as Ghosh et al.'s~\cite{Ghosh2009}, and Nissim and Brenner's~\cite{Nissim:10:FOCS}.}
We then scale-up the analysis using \emph{joint} algebraic properties of Kronecker products and \QIF.
%
%

Our main contributions are:
\Mf{Mario to rework on contributions once the other sections are reorganized.}

\begin{itemize}
%
	\item[$\cdot$] We introduce and explain the importance of \QIF's \emph{\textbf{refinement} (partial) order}
	(\Sec{s1150X})
	as a tool for rigorous understanding of the behaviour of utility measurements.
	\item[$\cdot$] We provide a novel model for the utility and privacy of statistics-release pipelines as in Fig.~\ref{fig:post-processing-pipelineX}
	(\Sec{s1008X}),
	by \textbf{combining} \QIF\ with Kronecker products: in particular we show how the Kronecker and \QIF\  algebra provides a smooth integration of \QIF's refinement order.
	(\Sec{s1009}).
	\item[$\cdot$] We use the resulting \QIF+Kronecker framework to study the curious phenomenon of (in)-\textbf{stability}, mentioned above;
	\item[$\cdot$] We provide both positive
	(``it's stable'': \Sec{s0914}, \Sec{s1347X})
	and negative
	(``it's not stable'': \Sec{s0915X}, \Sec{s1215X})
	and advisory
	(``it looks stable, but it might not be generally'': \Sec{sec:argmaxX})
	\textbf{recommendations} for data-release and --analysis practices.
\end{itemize}

\Mf{Plan of the paper?}

\section{Technical Background}

\renewcommand{\LFU}{\LF}
\newcommand\LFG {\ensuremath{{\ell}}}

We focus on perturbation mechanisms satisfying differential privacy, and on Bayesian utility measures adopted in previous works.
This section provides background on those, as well as existing results from \QIF\ essential for our study of stability.

\subsection{Quantitative Information Flow and the Refinement of Mechanisms}
\Label{s1150X}

Quantitative Information Flow (\QIF) is a mathematical framework for measuring information leaks from systems that are modelled as probabilistic channels. An important characteristic of \QIF\ is that its information-leakage measurements correspond to actual specific adversarial threats: the leakage of a system (wrt.\ a threat) is measured as the expected gain (or loss) of a Bayesian adversary whose (known) goals are modelled using a loss function, and whose prior knowledge is modelled using a Bayesian prior over secrets. We now detail the main tools from \QIF\ which we employ, and their important connections with differential privacy and the Bayesian model of utility.

\subsubsection{Channels and uncertainty}

In \QIF, a mechanism $M$ is an information-theoretic channel
\MStuff{of type $\calx{\to}\Dist{\caly}$,}
taking an input $x{ \in}\calx$ and outputting an ``observation'' $y{\in}\caly$ according to some distribution $\delta$ say in $\Dist{\caly}$, where in general \MStuff{$\Dist{\cals}$ is the set of
 	all probability distributions over a set $\cals$.} \Mf{Mario to do: Come back here later and add every other notation used to denote the type of a mechanism.}
 In the discrete case, such channels are $\XX{\times}\YY$ matrices whose row-$x$, column-$y$ element $M_{x,y}$ is the probability that input $x$ will produce output $y$. The $x$-th row $M_{x,-}$ is in that case a discrete distribution $\delta$ in $\Dist{\caly}$ as mentioned above.


\QIF\ assumes that adversaries are Bayesian: they are equipped with a prior $\pi{\In}\Dist{\calx}$ over secrets $\calx$ and use their knowledge of the channel/mechanism $M{\In}\calx{\to}\Dist{\caly}$ to maximise the advantage given by an observation of the mechanism's output.
This is modelled in full generality using the ``$g$-leakage framework'' of gain functions \cite{Alvim20:Book}. Here we use a dual formulation in terms of loss functions.\,%
\footnote{Duality for all of the results relevant to this paper has been shown in \cite{Alvim20:Book}.}

We begin with a general definition of loss functions.

\begin{Definition}[Loss function]\Label{d1357}
Given an input type \XX, a \emph{loss function} \LF\ operating on mechanisms $M$ with that input type is a non-negative real-valued function on $\XX{\times}\WW$, for some \emph{set of actions} \WW, so that $\LF(x,w)$ is the loss to the adversary should they take action $w{\in}\WW$ when the (hidden) input turns out to have been $x$.
\end{Definition}

For loss function $\ell$, the (prior) uncertainty wrt.\ the prior $\pi$  is the minimum loss to the adversary over all possible actions, defined $U_\ell(\pi) \Defs \min_{w{\In}\calw} \sum_{x{\In}\calx} \pi_x \ell(w,x)$; it  quantifies the adversary's success in the scenario defined by $\pi$ and $\ell$,
\MStuff{where $\pi_{x}$ is the probability $\pi$ assigns to element $x$.}
The \emph{(posterior) uncertainty} is then the adversary's \emph{expected} loss after observing $y$:



\begin{Definition}[Posterior uncertainty]\Label{d1418X}
For mechanism $M$ of input type \XX\ to \YY , loss function \LFU, and prior distribution $\pi$ on \XX, the \emph{posterior uncertainty} $U_\LFU$ of $M$ with respect to prior $\pi$ is
\[
	\Util{\LFU}{\pi}{M} \DefsR \sum_{y:\YY}~\min_{w:\WW}~\sum_{x:\XX}~\pi_x\cdot M_{x,y}\cdot\LFU(x,w)\Q.
\]
\Nf{Changed this to round bracket/usual function call syntax. \MStuff{Notice that, by doing that, we ended up with the same notation for $\ell$-uncertainty in Def.~\ref{d1418X} and Ghosh et al.'s measure in \EQn{eq:ghosh}. Since they end up being equivalent as per Lemma~\ref{l1333}, this may be an acceptable overloading of notation.}}%
That is, for each possible observation $y$ from the mechanism $M$, the (rational) adversary is assumed to choose an action $w$ in \WW\ that minimises the expected loss given by \LFU, i.e.\ averaged over both the prior $\pi$ and the probabilistic choices made in $M$.
\end{Definition}



\subsubsection{Refinement}

``Refinement'' of mechanisms is the principal theoretical tool that \QIF\ contributes to our study of privacy, utility, and stability. 

\begin{Definition}[\QIF-refinement \cite{Alvim20:Book}]\Label{d1526X}
One mechanism $M$ is said to be  \emph{refined by}
another $M'$, written $M\,{\ChanRef}\,M'$, just when for all priors $\pi$ and all loss functions \LFU,
the adversary's (expected) loss from using $M$ is no more than their loss from using $M'$, i.e.\ \\
\begin{minipage}[m]{0.9\linewidth}
\centering
\begin{align*}
	M\CR M' \quad\textit{ iff }\quad \Util{\LFU}{\pi}{M} \leq \Util{\LFU}{\pi}{M'}\quad \textit{ for all } \pi, \ell~.\qedhere
\end{align*}
\end{minipage}
\hfill
\begin{minipage}[b]{0.08\linewidth}
	\qedhere
\end{minipage}
\end{Definition}

The above definition characterises refinement in terms of safety against adversarial threats:\,%
\footnote{Note that in the \QIF\ literature, refinement is often equivalently defined in terms of gain functions rather than loss functions.}
it tells us that $M$ is refined by $M'$ just when $M'$ is safer than $M$ against \emph{any} Bayesian adversary modelled using a loss function and a prior; alternatively, every Bayesian adversary prefers $M$ to $M'$\MStuff{, as the first is always guaranteed to be ``at least as  leaky'' as the second (i.e. for every $\pi, \ell$). }

\Def{d1526X} does not however tell us how to verify when refinement holds; for this we use the Coriaceous theorem (below) which provides a complementary \emph{structural} characterisation of refinement.

\begin{Theorem}[{Coriaceous~\cite{McIver:2014aa}}]\Label{t1516X}
Mechanism $M$ is refined by $M'$ just when there is a stochastic-matrix ``witness'' $W$ such that
$M{\MM}W{=}M'$ where {\rm($\MM$)} here is matrix multiplication.\,%
\footnote{This is a soundness/completeness result if $\LFU/\pi$ is regarded as testing $M$ \cite[Sec.\,9.5]{McIver:2014aa}: if there's a $W$, then any $\LFU/\pi$ -test must go the right way; if every $\LFU/\pi$ -test goes the right way, then there's a $W$.}
Note that all three matrices must be stochastic for refinement to hold.
\end{Theorem}

\NStuff{Finally we recall~\cite[Sec. 4.6.2]{Alvim20:Book}
   that if $C$ is a channel and $P$ is a post-processing of the output from $C$, then the result of applying $C$ followed by $P$ (i.e. their sequential composition) is exactly $C{\MM}P$ where {\rm($\MM$)} is matrix multiplication. Thus, with \Thm{t1516X}, we see that post-processing corresponds to refinement: this is the ``data-processing inequality'', or \DPI.}

%

\subsection{Differential privacy}\Label{s1423X}

Differential privacy is an influential approach, introduced by Dwork et al., to measuring the indistinguishability of secrets in adjacent datasets~\cite{Dwork2006}.
\MStuff{The notion  has later been extended to other domains,
including those formed by individual records
themselves~\cite{doi:10.1137/090756090}, or subject to metrics modelling general notions of indistinguishability~\cite{chatzikokolakis2013broadening,DBLP:phd/hal/Fernandes21}.}
Here we adopt differential privacy as our measure of the privacy afforded by a probabilistic mechanism.

A \emph{mechanism} $M$ taking input data \XX\ to outputs \YY\ (as above) is said to satisfy \emph{\VE-differential-privacy}, abbreviated \EDP, if the distributions $\Delta_{1,2}$ on \YY\ that are produced respectively by $M(x_1)$ and $M(x_2)$, i.e.\ by the mechanism $M$'s acting on ``adjacent'' elements $x_{1,2}$ of \XX, 
are at least \VE-close in the sense that for all subsets $Y$ of \YY\ we have
\begin{equation}\label{e1048-oldX}
	\left|\;\ln\;\nicefrac{\Delta_1(Y)}{\Delta_2(Y)}\;\right|
	\WIDE{\leq}
	\varepsilon \Q.
\end{equation}
The smaller the $\varepsilon$, the more private the mechanism $M$ is said to be.

The definition of ``adjacency'' depends on the chosen \XX\  \MStuff{and on the desired notion of indistinguishability.
In the original formulation of \EDP\ \cite{Dwork2006}, also known as  \emph{central differential privacy}, the data \XX\ are datasets of rows of data (about e.g.\ individual people), and two datasets $x_{1,2}$ are said to be adjacent just when they differ in the presence or absence of a single row. 
In \emph{local differential privacy}~\cite{doi:10.1137/090756090}, the data \XX\ are single values (or rows), and every pair $x_{1,2}$ of values (rows) are considered adjacent --- the idea being that a single data subject's value (row) should be indistinguishable from any other value (row) this individual might have contributed.}

If the distributions $\Delta$ on \YY\ produced by mechanism $M$ acting on some $x$ are known to be discrete, then \Def{e1048-oldX} can be simplified to taking the ratio of ``the probabilities $p_{1,2}$ assigned to $y$ by $M(x_{1,2})$ for all $y$ in \YY'',
\Mf{What's the role of \YY'' (double prime) here?}
 i.e.\ to individual elements rather than the (more general, perhaps only measurable) subsets of \YY. We assume discrete distributions here.

 \Nf{Removed text about satisfies vs realises. Lets not use different words here. Realises and satisfies are too easy to forget. Just say what we mean.}

Both the input data type \XX\ and its adjacency relation contribute to the differential privacy \VE\ of a mechanism $M$, but (we recall) the prior distribution on \XX\ does not. In that sense, some say that differential privacy is independent of prior knowledge. \Mf{Catuscia would have a heart attack by reading this slogan :-) I changed the phrasing to ``is said to be'', because in fact it isn't actually (except for in a very narrow sense). \NStuff{Good point. Perhaps we can footnote it. We could also say ``is often described as''.}}


\subsubsection{Refinement's relation to \EDP}
\Label{s1144X}

There is a strong 
relationship between refinement of mechanisms and the values of \VE\ they realise in \EDP\J, \NStuff{ as the following lemma shows},
\QIF-refinement preserves differential privacy:
\begin{lemma}[$\VE$-preserving \cite{DBLP:conf/aplas/McIverM19,Chatzi:2019}]\Label{l1430}
If $M, M'$ are mechanisms with $M\CR M'$, and
$M$ satisfies \EDP\ for some \VE, then so does $M'$.
\end{lemma}

\NStuff{In other words, the refinement order on mechanisms is (strictly) stronger than the \VE-ordering of differential privacy.}  In fact \Lem{l1430} captures, although abstractly, the well-known  principle that privacy is not reduced through post-processing: that post-processing induces refinement (the \DPI\,).

In general, \Lem{l1430} is only one-way. 
In some cases however it holds in both directions, in particular when the mechanisms $M, M'$ belong to the same ``family'' \cite{Chatzi:2019}. We call that property of sets (of mechanisms) ``refinement-preserving'':

{
\begin{definition}[refinement-preserving]\Label{d1322}
A set \FFF\ of mechanisms is said to be \emph{refinement preserving} if for any $M, M' \in \FFF$ satisfying $\VE, \VE'$-DP respectively (where $\VE, \VE'$ are the minimal such values), and $\VE'{ \leq}\VE$, it holds that $M{\ChanRef}M'$.
\end{definition}

Refinement-preservation for a set $\FFF$ means
that --within that set \FFF\,-- \Lem{l1430} holds in \emph{both} directions: that is, given an $M$ in \FFF, choosing a more \EDP-private mechanism $M'$ from the same $\FFF$ guarantees that
utility will decrease.  This crucial idea simplifies the process of finding a good trade-off between privacy and utility.
} 

Unfortunately there are many examples of sets of mechanisms which \emph{do not} satisfy \Def{d1322}, most notably those identified by Ghosh et al. \cite{Ghosh2009} and Brenner and Nissim \cite{Nissim:10:FOCS}.
%
Interestingly there are also sets of mechanisms that \emph{do} satisfy \Def{d1322} --- for example the notion of ``families'' of mechanisms as mentioned above, loosely defined in the literature to mean mechanisms that are constructed using the same noise distribution, so that different mechanisms from the same family differ only in how that noise distribution is parametrised.
Two important families for which \Def{d1322} have been shown to hold (see \cite{Chatzi:2019}) 
are the family of random-response mechanisms \MStuff{(Def.~\ref{d1451X} ahead)},
which perturb the ``true'' response amongst $K{-}1$ others,
and the family of geometric mechanisms
\MStuff{(Def.~\ref{d2351X} ahead)},
which perturb an integer value according to the geometric distribution.
~\Nf{Can we be consistent with names here. I think we should use ``random response mechanism'' and ``geometric mechanism''. Avoid ``distribution''.}



\subsection{Measures of utility, \\and their relation to refinement}\Label{s1425X}

In addition to the privacy-preserving properties of a noisy mechanism, an important measure is its utility, how
\emph{useful} it is to a data analyst.
Achieving an optimal privacy-utility balance is a key goal in the design of privacy mechanisms and in the tuning of $\VE$ alluded to above. The measure of utility we adopt here is a general Bayesian model proposed by  Ghosh et al. \cite{Ghosh2009}.
 In the original Ghosh et al. setting, a consumer (analyst) is identified by a prior $\pi$ over inputs \XX\ and a loss function \LFG\ which gives the cost $\LFG(x', x)$ to the consumer if she guesses
the value $x'$ when the true value of the input was $x$.\,%
\footnote{In our terms, Ghosh et al.'s loss function \LFG\ has $\WW\,{=}\,\XX$.}

The utility of a mechanism $M$ taking inputs \XX\ to observations \YY\ is given by the minimum \emph{expected value} of $\LFG(r(y), x)$ where the minimisation is taken over all possible ``remappings'' $r$ from observations to inputs, i.e.
\begin{equation}\label{eq:ghosh}
   U_\ell(\pi, M) \Wide\Defs \min_r \sum_{x \in \XX} \pi_x \sum_{y \in \caly} M_{x,y} \cdot \LFG(r(y), x) ~,
\end{equation}
\noindent where $M_{x, y}$ is the conditional probability $Pr(Y=y\mid X=x)$.
\Gf{Is the right side of \ref{eq:ghosh} correct? Both summations are over the query results in \cite{Ghosh2009}.}
%
That is, the (Bayesian) consumer uses their prior knowledge and the known probabilistic perturbations caused by the mechanism $M$ to maximise their expected utility (minimise their expected loss \LFG\ ).

The following lemma shows that our use of \LFU\ in \Def{d1418X} in fact subsumes Ghosh et al.'s \LFG-formulation in \Eqn{eq:ghosh} above: Ghosh's remapping function can be treated as a choice of action $w$ from \WW\ \cite{DBLP:conf/csfw/FernandesMPD22}.
\begin{lemma}[Optimal utility \cite{DBLP:conf/csfw/FernandesMPD22}]\label{l1333}
	The expected \LFG-loss in \EQn{eq:ghosh}  is equivalent to the posterior \LFU-uncertainty of \Def{d1418X}.
\end{lemma}

\noindent Thus here-on we use \Def{d1418X} as our general formulation for expected utility.

\Thm{t1516X} and \Def{d1418X} together give a direct relationship between refinement and utility -- that refinement $M \CR M'$ occurs exactly when the output of $M$ gives more utility than the output of $M'$, as measured by the expected gain/loss of a Bayesian consumer. This relationship has previously been explored in the context of optimality \cite{DBLP:conf/csfw/FernandesMPD22}.

\section{Two-stage data-release pipelines, and the notion of stability}
\Label{s1008X}

\Mf{Natasha, \Nz, I think this Section~\ref{s1008X}
introduces the meat of the paper, and its goal should be to convince the reader now (before digging into complicated theory), that:
	\begin{itemize}
		\item \textbf{Point 1 -- Our model:} QIF can model the relevant, real-world pipelines (including the composition of perturbation and post-processing steps); and
		\item \textbf{Point 2 -- Stability:} the concept of stability matters.
	\end{itemize}
	If we can do that here, the rest of the paper will flow much more easily.
	In my view Point 1 (Our model) expands \Sec{s1150X} with
	the QIF background that is specific to our paper,
	and is part of our proposed model, whereas
	Point 2 (Stability) introduces the really novel stuff.
	(We can think of perhaps splitting the 2 points above into their own sections in the future, but for the time being I think it makes sense for them to be just subsections of this current section.)
	For the explanation of our pipelines, I think we can have subsections dedicated to present clearly what typical perturbers are (e.g., RR and Truncaged Geometric), \textbf{including their definitions and algorithms (which are standard anyway!)}, and what typical post-processors are (e.g., counting, average, max, etc.).
	We also have to present how QIF models their composition.
	It would be nice to collect all these (more or less standard) definitions in one place and keep them separated from the deep, theoretical parts of the paper (specially because some mechanisms/queries are now defined in the middle of a non-background section, and used all over, and it makes it harder for the reader to know where concepts were introduced).
	Moreover, by introducing all mechanisms/queries here, we would have \textbf{a centralized place (perhaps even a table of notation?) to
	define all mnemonics (like $P$, $S$, $T$, $M$, etc.}) which now are spread all over the paper and are hard to follow.
	\textbf{This would make the life of the reader much easier.}
	For the reasons above, I was thinking of a structure for this section more or less like the following:
\begin{enumerate}
	\item \textbf{Point 1 -- Our model:} A detailed description of our pipeline in two stages \textbf{(with an updated, beautiful picture with mnemonics)}, and an explanation of how cascading captures the composition of the functions representing these stages (not every reader knows that matrix multiplication captures post-processing!).
	Include examples of:
	\begin{enumerate}
		\item what the perturber can be: define RR and Truncated Geometric, with their typical algorithms \textbf{(for the case of 1 individual only; we can leave multiple individuals -- which will be modelled via $N$-way Kroneckering anyway, for later in the paper)}.
		\item what the post-processor can be: define what we use: counting, max, average, etc.
		(We do not necessarily need to put the matrices here, but only the description of the query. The matrices can be introduced as needed.)
	\end{enumerate}
	\item \textbf{Point 2 -- Stability:} A definition of what a ``family'' (or whatever concept replaces it) is.
	A Definition of stability and why it matters.
\end{enumerate}
}

We now formalise the concept of stability outlined in the introduction, and give some concrete examples of perturbation mechanisms and post-processors, starting with pipelines $M$ whose structure is as in \Fig{fig:post-processing-pipelineX}. It turns out that its two components --namely the perturbation stage and the post-processing stage-- can both be modelled using channels, enabling a convenient formalisation of the whole pipeline as standard matrix multiplication. And that can then be studied using the techniques of \Sec{s1150X}.

\begin{definition}[Pipeline model]\Label{d1523}
A \emph{two-stage pipeline structure} is a probabilistic-perturbation step represented by a stochastic matrix $C\,{\in}\,\Matrix{\XX}{\YY}$ that takes original data to perturbed data followed by a post-processing step represented by a deterministic matrix $P\,{\in}\,\Matrix{\YY}{\ZZ}$ that takes perturbed data as input and combines it to produce a released output (such as a statistic). The whole pipeline is then $M = C\cdot P$.

The \emph{utility} of the pipeline $M$,  for a Bayesian analyst who has prior $\pi\in \Dist\XX$ and loss function $\ell$, is then $\Util{\LFU}{\pi}{C\cdot P}$, where $\XX$ represents the set of potential values of the raw data.
\end{definition}

Examples of popular loss functions for measuring accuracy are:
\begin{eqnarray}
 \textit{Bayes' risk}:&  {\sf br}(w, x)\Defs 0~\textit{if}~(w{=}x)~\textit{else}~ 1~; \label{BR}\\
 \textit{Linear error}:& {\sf le}(w, x)\Defs |w{-}x|~;\\
 \textit{Mean-squared error}:& {\sf mse}(w, x)\Defs |w{-}x|^2~.
\end{eqnarray}
Bayes' risk models an analyst intent on guessing the whole secret, whereas linear and mean-squared errors model an analyst seeking only to obtain a good approximate value.  Notice that the utility applies to the whole pipeline, since the analyst takes a measurement only after the post-processing step has completed.

Examples of perturbers include those we have already mentioned, such as those based on the geometric distribution, but can also include any probability distribution applied to the data. Finally, examples of post-processors include the computation of any value from the perturbed data, such as  counting records that satisfy a property,  determining totals or averages. The popular $\ArgMax$, which returns the mode value  of a histogram, is also an example of a post-processor. Utility, in all those cases, then provides a measure for how accurate those computed values turn out to be. Note  that the architecture described by \Def{d1523} frequently appears in privacy-preserving libraries, albeit in more complex implementations; and we provide simple code-based examples of them below. \Def{d1523}'s  concept of perturbing then post-processing is used, in part,  by  the very well-known Rappor \cite{Rappor},  and the noisy version of $\ArgMax$ (treated below) as well as other complex programs detailed in \cite{Dwork2013}.   Using  those code libraries  usually requires the  definition of a privacy parameter $\VE$ which can then be ``tuned'' to obtain the best accuracy.
%
%
\Mf{If the reader is not a QIF person, we need to show (or at least argue convincingly) that post-processing via cascading is the equivalent of composing functions.}
\Af{Why isn't it obvious?}
\NStuff{In such pipelines we are interested in the effect of the post-processing step on the overall utility of the system; and a useful property to draw upon during tuning would be that an increase in the $\VE$ would lead to an increase in accuracy. We use the \QIF\ notion of refinement to define this ``stable'' pipeline behaviour.}



\begin{definition}[Stability]\Label{d1632}
Let \CCC\ be a set of mechanisms  in \Matrix{\XX}{\YY}, and let \PPP\ be a set of post-processors in \Matrix{\YY}{\ZZ}.
We call the pair $(\CCC,\PPP)$ \emph{stable} just when for all $C,C'$ in \CCC\ and $P$ in \PPP\ we have that
\begin{equation}\label{e1516X}
	C\ChanRef C' \WIDERM{implies} C\MM P \,\ChanRef\, C'\MM P \Q.
\end{equation}
\end{definition}

\NStuff{Stability, then, tells us when post-processing preserves refinement, or in other words when it preserves the utility order of the mechanisms: if $C$ is more useful than $C'$, then so will be the corresponding pipelines $C\MM P$ and  $C'\MM P$.  Unfortunately this property  does \emph{not} hold in practice in many widely-used scenarios. }

\Nf{I am not sure what stability of pipelines is trying to say, removing it for now.}
\NStuff{However, \Def{d1632} is very strong: it holds for \emph{all} loss functions measuring utility. In practice we might be interested in stability for only a restricted set of loss functions --- and so we define a weaker notion of stability called \LF-stability:}

\begin{definition}[\LF-stability]\Label{d1537}
Let \CCC\ be a set of mechanisms, \LF\ a loss function, and $P$ a post-processor.
We say that the pipeline $\CCC\MM P$ is \emph{\LF-stable} just when for any $C,C'\,{\in}\,\CCC$ with $C\CR C'$ we have
\[
	\Util{\LF}{\pi}{\,C{\MM}P\,} \Wide{\leq} 	\Util{\LF}{\pi}{\,C'{\MM}P\,}
\]
for any prior $\pi$.
\end{definition}
Note that from \Def{d1322}, if \CCC\ is refinement-preserving then refinement $C\CR C'$ is induced by decreasing the \VE\ of $C$.

We shall see that in some scenarios, a pipeline $C\MM P$ might be \LF-stable (i.e.\ for some \LF\,), even when it is not (universally) stable (i.e.\ for \emph{all} \LF\,).

%

\subsection{Examples of perturbing mechanisms \\and post-processors}

Up until this point we have --abstractly-- described data-release pipelines consisting of perturbing mechanisms and post-processors. In this section we introduce specific examples of mechanisms and post-processors for which we will prove results on stability in the coming sections.


\subsubsection{Perturbers: Randomised response and Geometric mechanisms}


\NStuff{We recall two popular families of mechanisms from the literature: the randomised-response family and the geometric family of perturbation mechanisms. Both of those families have been shown to be refinement-preserving (\Def{d1322}).}

\NStuff{
\begin{Definition}[Random-response mechanism]\Label{d1451X}
	The \emph{random-response mechanism} on $K$ values, written $\RR^K$, is described by a channel
	whose value at row $r$ and column $c$ is given by
	\[
	\begin{array}{lllp{10em}}
	(\RR^K)_{r,c}
	& = & p{+}\NF{\Pn}{K} & if~\mbox{~$r{=}c$} \\
	&& \nicefrac{\Pn}{K} & otherwise\Q,
	\end{array}
	\]\Mf{Why not use the usual ``cases'' environment here?}
	where $p$ satisfies $Kp/\Pn+1 = \exp(\VE)$, and $\RR^K$ satisfies $\VE$-DP.
	\end{Definition}
}

\MStuff{As a computer program, random response over $K$ choices is implemented by \Alg{a1406X}, where for convenience we have set $\KK{=}\{1..K\}$. }
The special case in matrix form when $K{=}4$ is given in \Fig{f1026X}.
\Nf{Do we need this algorithm? \AStuff{Yes .. in Alg 2}}

\begin{algorithm}
	\begin{verbatim}
  # Random response: Realises eps-DP
  #                  for exp(eps) = 1+(Kp/(1-p)).

  def RR(p,K,x): # The secret x is in 1..K.
           y = x      # With prob p tell the truth, else
      [p]  y<-(1..K)  # With prob 1-p choose y uniformly.
      return y
	\end{verbatim}
	\caption{~Implementing a one-respondent $p$-truth random \\\hspace*{14ex}response over $K$ choices}\label{a1515}
	\label{a1406X}
\end{algorithm}

\MStuff{
\begin{Definition}[Truncated geometric mechanism]\Label{d2351X}
The (truncated) $\alpha$-geometric mechanism is given by:
\[
	(\GD[\alpha])_{r,c} \Wide{=} \NF{1-\alpha}{1 + \alpha} \,\cdot\, \alpha^{|r-c|}\Q,
\]
For $\VE=-\ln\alpha$, the mechanism satisfies $\VE$-DP, and we write \GD\ in that case.
\Cf{Why do we use $\|-\|$ instead of absolute value $|-|$\,?}
\end{Definition}
}

The truncated geometric mechanism is popular when the data values are numeric; it is often used in the ``central model'' for differential privacy, but can also be used in implementations for noisy \ArgMax\  and   has been used in local differential privacy \cite{10.1145/3264820.3264827}.

\subsubsection{Post-processors}

\NStuff{Post-processors can be thought of as statistical aggregators, such as the well-known counting and sum ``queries''. As channels, they are deterministic in the sense that they consist only of  0's and 1's, and they have the effect of `collecting' or `grouping' columns of the perturbing mechanism. For example, a counting post-processor applied after a Geometric mechanism might count the number of individuals whose marital status is set to `married'. In that case, the post-processor merges columns of the Geometric mechanism producing the same number of `married' individuals.
}

\NStuff{
\begin{definition}[Deterministic post-processor]\Label{d1012}
A \emph{deterministic post-processor} is a channel matrix \Matrix{\YY}{\ZZ}, such that $P_{y,z}$ is either $0$ or $1$, with exactly one $1$ in each row, and for each column $z\,{\in}\,\ZZ$, there is some $y\,{\in}\,\YY$ such that  $P_{y,z}=1$.

The set of deterministic post-processors is called \PPP.
\end{definition}
Each $P$ thus represents a total (but not necessarily injective) function from \YY\ to \ZZ. An interesting property of deterministic post-processors, and one that will be useful in our development below, is that they have left inverses, i.e. for each $P$ there is a matrix which we write \PI\ and which satisfies
\begin{equation}\label{e1001}
\PI\MM P=\Id \Q,
\end{equation}
where \Id\ is the identity in \Matrix{\ZZ}{\ZZ}.
Note that left-inverses (when they exist) are generally not unique --- but they are stochastic.}

\NStuff{In the next sections we will study the property of stability on various data-release pipelines consisting of perturbation mechanisms followed by post-processors. Recalling that stability is a property of the utility of the overall pipeline, we will see that stability is not at all assured even when the perturbation mechanisms themselves maintain a privacy-utility balance via $\DP$-refinement.}



\section{Guaranteed stability}
\subsection{Random response with one respondent}
\Label{s0914}


\NStuff{We consider a single respondent who uses a random-response mechanism $R=\RR^\XX$ of type \Matrix{\XX}{\XX} to conceal his secret value $x{\in}\XX$. We also consider a second possible perturber $R'=\RR[\VE']^\XX$ with $\VE'{\leq}\VE$, of the same type.}
Since the set of random-response perturbers is a family \cite{Chatzi:2019}, we have $R\CR R'$.

\begin{figure}
\[
    \raisebox{-3.3ex}{Input \XX\ to $\RR^4$}
    \hspace{1em}
    \begin{array}{c|cccc}
        \multicolumn{1}{c}{~}
        	&	\multicolumn{4}{c}{\textrm{Output \YY\ from $\RR^4$}} \\[1ex]
        	&	1 				&	2					&	3						& 	4 \\\hline\\[-1.5ex]
        1	&	p{+}\NF{\Pn}{4}	&	\NF{\Pn}{4}			&	\NF{\Pn}{4}				&	\NF{\Pn}{4} \\
        2	&	\NF{\Pn}{4}		&	p{+}\NF{\Pn}{4}		&	\NF{\Pn}{4}				&	\NF{\Pn}{4} \\
        3	&	\NF{\Pn}{4}		&	\NF{\Pn}{4}			&	p{+}\NF{\Pn}{4}			&	\NF{\Pn}{4} \\
        4	&	\NF{\Pn}{4}		&	\NF{\Pn}{4}			&	\NF{\Pn}{4}				&	p{+}\NF{\Pn}{4}
	\end{array}
\]
In the matrix above $p$ is set as in \Def{d1451X}.
\caption{Matrix representation of $\RR^4$}\label{f1026X}
\end{figure}

\NStuff{We also note the special property of random-response mechanisms --call that set \RRRR-- that if $R \ChanRef R'$ then the refinement witness $W$ satisfying $R{\cdot}W = R'$ is itself in \RRRR,
that is $R\MM R''=R'$ for some $R''=\RR[\VE'']$.} 
\NStuff{We now prove that random-response mechanisms are stable for any post-processor:
}
that is $(\RRRR,\PPP)$ is stable. We begin with a technical lemma.

\begin{lemma}\Label{l1601X}
Let deterministic post-processor $P$ be in \PPP, and let \PI\ be some left-inverse of it. For any member $R''$ of family \RRRR\ we have
\[
	P\MM\PI\MM R''\MM P \Wide{=} R''\MM P \Q.
\]
\begin{proof}
From \Def{d1451X} we have that $R''$ can be written as $a\Id+b\Uf$, for some $0{\leq}a,b{\leq}1$ and \Uf\ the
everywhere \NF{1}{K} matrix in \Matrix{\XX}{\YY}.
We then reason
\begin{Reason}
\Step{}{P\MM\PI\MM R''\MM P}
\StepR{$=$}{$R''=a\Id+b\Uf$}{
	P\MM\PI\MM(a\Id+b\Uf)\MM P}
\StepR{$=$}{distribution}{aP\MM\PI\MM\Id\MM P ~+~ bP\MM\PI\MM\Uf\MM P}
\StepR{$=$}{\,$\PI\MM\Id\MM P = \Id$\J}{
	aP\MM\Id ~+~ bP\MM\PI\MM\Uf\MM P}
\StepR{$=$}{$P\MM\Id=\Id\MM P$ and $P\MM\PI\MM\Uf=\Uf\,(\dagger)$\J}{
	a\Id\MM P ~+~ b\Uf\MM P}
\StepR{$=$}{distribution}{(a\Id + b\Uf)\MM P}
\StepR{$=$}{$a\Id+b\Uf=R''$}{R''\MM P\Q.}
\end{Reason}
The $(\dagger)$-step uses that all of the matrix $\Uf$'s elements are equal.
\end{proof}
\end{lemma}

\NStuff{We now give the main result of this section.}

\NStuff{
\begin{theorem}[Stability of random response]\label{thm:rr_stability}
$(\RRRR,\PPP)$ is stable for the entire set \PPP\ of deterministic post-processors (\Def{d1012}).

\begin{proof}
We must show that for any $R\CR R' \in \RRRR$ and $P$ any deterministic post-processor we have $R{\cdot}P \ChanRef R'{\cdot}P$.

Let \PI\ be a left-inverse of $P$. 
We reason as follows:
\begin{Reason}
\Step{}{R'\MM P}
\StepR{$=$}{above, some $R''$}{R\MM R''\quad\MM P}
\StepR{$=$}{associativity}{R\MM\quad R''\!\MM P}
\Space
\WideStepR{$=$}{\Lem{l1601X}, using the fact that the witness $R''$ is some $\RR[\VE'']^K$}{R\MM\quad P\MM\PI\MM R''\!\MM P}
\Space
\StepR{$=$}{associativity}{R\MM P\quad\MM(\PI\MM R''\!\MM P) \Q,}
\end{Reason}
whence $R\MM P~\CR~R'\!\MM P$, with (stochastic) witness $\PI\MM R''\!\MM P$.
\end{proof}
\end{theorem}
}

\Nf{Not sure what is the purpose of this section, removing for now.}

\subsection{Multiple respondents via Kronecker}\Label{s1322}

The above section dealt with the case of a single respondent. In more general cases however, there are \emph{many} respondents --- with each one independently executing the same kind of 1-respondent perturbation as above. For this analysis we will use a construction called the ``Kronecker product'' (\Sec{s1057X}) to build many-respondent perturber-matrices from single-respondent ones.

The procedure works with any perturber, but we will use random-response as an example.
Recall from \Alg{a1406X} that random-response uses a probability-\verb+p+-coin flip (denoted \verb+[p]+) to choose between its two adjacent statements.
If we call \Alg{a1406X} for each of the $N$ respondents' data, record the results and then post-process by counting to form a complete pipeline, we get \Alg{alg:nfold_rrX}.

\begin{algorithm}
\begin{verbatim}
# The original responses are in X[0:N],
# a sequence of choices in (say) 1..K.
for n in range(N): # Randomise the X's to produce the Y's.
    Y[n]= RR(p,K,X[n])
# The RR_p -perturbed choices are now in Y[0:N].

count= 0
for n in range(N): # Count the Yes responses.
    count= (count+1 if Y[n]==Yes else count)
# count is the number of Yes's among the perturbed Y's.

print(count)
\end{verbatim}
\caption{A random-response over $N$ respondents each with $K$ choices, implemented by \Alg{a1406X}
followed by a counting query.}
\label{alg:nfold_rrX}
\end{algorithm}



%
%
%

If we take that \Alg{alg:nfold_rrX} and specialise to a random-response channel $R=\RR[p=2]=\RR[\ln3]$,
a choice-set $\KK{=}\{\Yes,\No\}$ so that $K{=}2$, and then consider only $N{=}2$ respondents,
%
%
%
%
%
we would represent the perturbation for a single respondent by a $2{\times}2$ matrix where the rows are inputs \XX\ and the columns are outputs \YY:
\begin{equation}\label{eqn:rr-channelX}
\begin{array}{ c | cc  }
    R & \Yes & \No \\
    \hline
    \Yes & 3/4 & 1/4 \\
    \No & 1/4 & 3/4 \\
\end{array} \Q.
\end{equation}

\noindent It outputs the true answer with probability $3/4=1/2+\NF{1}{2}{\times}\NF{1}{2}$; it ``lies'' with probability $1/4=\NF{1}{2}{\times}\NF{1}{2}$.

\NStuff{Considering now \emph{two} participants each applying $R$ to their secret input, we construct the $4{\times}4$ channel $\nfold{2}{R}$ corresponding to the probabilities of observing ordered pairs of outputs given the input as an ordered pair:
}
\begin{equation}\label{eqn:r2_channelX}\small
  \begin{array}{c | c c c c }
 $\nfold{2}{R}$  & (\Yes, \Yes) & (\Yes, \No) & (\No, \Yes) & (\No, \No) \\
  \hline
  (\Yes, \Yes) & 3/4\times3/4 & 3/4 \times 1/4 & 1/4 \times 3/4 & 1/4 \times 1/4 \\
  (\Yes, \No) & 3/4 \times 1/4 & 3/4 \times 3/4 & 1/4 \times 1/4 & 1/4 \times 3/4 \\
  (\No, \Yes) & 1/4 \times 3/4 & 1/4 \times 1/4 & 3/4 \times 3/4 & 3/4 \times 1/4 \\
  (\No, \No) & 1/4 \times 1/4 & 1/4 \times 3/4 & 3/4 \times 1/4 & 3/4 \times 3/4
  \end{array}
\end{equation}
The output-pairs in $\YY=\{\Yes,\No\}^2$ (the column labels) are what are sent to the analyst's post-processor for counting. Calling that post-processor ``$T$'' (for ``coun\underline{T}ing'') suggests the definition
\begin{equation}\label{e1020X}
   \begin{array}{ c | c c c }
          T & 0 & 1 & 2 \\
          \hline
          (\Yes,\Yes) & 0 & 0 & 1 \\
          (\Yes,\No) & 0 & 1 & 0 \\
          (\No, \Yes) & 0 & 1 & 0 \\
          (\No, \No) & 1 & 0 & 0
         \end{array}
         \Q,
\end{equation}
and the combined pipeline $\nfold{2}{R}\MM T$ represents \Alg{alg:nfold_rrX} when $N{=}2$, and $K{=}2$. The variable \verb'count' in the program code becomes the column-label of $T$. (Note that $T$ is deterministic.)

We return to counting-via-matrices in \Sec{s1350T}.

\subsection{The Kronecker product}\Label{s1057X}

The matrix operation that produced $\nfold{2}{R}$ from $R$ above is called a \emph{Kronecker product}, which we write as a binary operator $\kron$.
Here is the definition of the Kronecker product in general:
\begin{definition}\Label{d1130X}
For real-valued matrix $A$ with rows in \XX\ and columns in \YY, which type we write \Matrix{\calx}{\caly}, and $A'$ of similar type \Matrix{\calx'}{\caly'}, the Kronecker product $A\kron A'$ of type \Matrix{\calx{\times}\calx'\,}{\,\caly{\times}\caly'} is defined
\[
(A\kron A')_{(x,x'),(y, y')} \Wide{=} A_{x,y}\times A'_{x',y'} \Q.
\]
We write $\nfold{N}{A}$ for the $N$-fold product of $A$ with itself: thus $\nfold{N}{A}$ is $\nfold{N{-}1}{A}\kron A$ and $\nfold{0}{A}$ is the $1{\times}1$ identity \Id.
\end{definition}

As an example of the above, we note that $A\,{\kron}\,A'$ is always defined (i.e.\ for matrices $A,A'$ of any shape), and is constructed by a systematic multiplication of every element of $A$ by every element of $A'$. For example, if $A$ has shape $m{\times}n$ and $A'$ has shape $p{\times}q$ then $A\,{\kron}\,A'$ has shape $mp{\times}nq$, as given here:
\[
     A \kron A' \Wide{=} \begin{bmatrix}
                        A_{1,1}\times A'_{1,1} & A_{1,1}\times A'_{1,2} & \ldots & A_{1,n}\times A'_{1,q} \\
                        A_{1,1}\times A'_{2,1} & A_{1,1}\times A'_{2,2} & \ldots & A_{1,n}\times A'_{2,q} \\
                        \vdots & \vdots & \vdots & \vdots \\
                        A_{m,1}\times A'_{p,1} & A_{m,1}\times A'_{p, 2} & \ldots & A_{m,n}\times A'_{p,q}
                        \end{bmatrix}
                        \Q.
\]
The algebraic properties of $\kron$ include that it is
\begin{description}
\item[\it Associative:\quad] $(A \kron B) \kron C = A \kron (B \kron C)$\Q,
\item[\it Bilinear:\quad] $A \kron (B + C) = A \kron B + A \kron C$ \\ and $(B + C) \kron A = B \kron A + C \kron A$\Q,
\item[\it Product-respecting:\quad] If $A\MM C$ and $B\MM D$ are defined \\ then $(A \kron B)\MM(C \kron D) = (A\MM C) \kron (B\MM D)$\quad{\rm , and}
\item[\it Invertible:\quad]  If $A, B$ are invertible, then so is $A \kron B$. \\ The inverse $(A \kron B)^{-1}$ is $A^{-1} \kron B^{-1}$\Q.
\end{description}
But $\kron$ is not commutative in general.\,%
\footnote{There are some examples where commutativity does hold e.g. $\nfold{N}{C}\kron C = C\kron \nfold{N}{C}$.}

The channel that is implemented by the first part of \Alg{alg:nfold_rrX} is thus the $N$-fold Kronecker product of those independent responses.

\subsection{Kronecker products preserve refinement}
\Label{s1009}

A surprising consequence of $\kron$'s algebraic properties is that it respects refinement
(\Sec{s1150X}): the \emph{product-respecting} property of Kronecker product, together with \Def{d1526X}, gives an immediate relationship between the two. We have

\begin{Lemma}\Label{l0907X}
Let $C \sqsubseteq C'$ be two channels with refinement witness matrix $W$, and recall that $\nfold{N}{(-)}$ is the $N$-fold Kronecker product of $(-)$ with itself.
%
Then $\nfold{N}{C} \sqsubseteq \nfold{N}{C'\,}$\,, with refinement witness $\nfold{N}{W}$.
\begin{Proof}
We show by induction on $N$ that $(\nfold{N}{C})\MM (\nfold{N}{W})= \nfold{N}{C'}\!$ (as in \Thm{t1516X}).

Observe first that whenever $C,C'$ are conformal (i.e.\ their multiplication is possible) then from \Def{d1130X} so are $\nfold{N}{C}$ and $\nfold{N}{C'}\!\!$.

For $N{=}1$ the equality holds because $W=\nfold{1}{W}$ is the witness for $\nfold{1}{C} \CR \nfold{1}{C'}$. Then the simple induction is
\begin{Reason}
\Step{}
{\nfold{N+1}{C}\,\MM\,\nfold{N+1}{W}}
\StepR{$=$}{Unfold recursion}
{(\nfold{N}{C}\, \kron C)\,\MM\,(\nfold{N}{W}\, \kron W)}
\StepR{$=$}{Respects matrix products}
{{(\nfold{N}{C}\,\MM\,\nfold{N}{W}) \kron (C\MM W)}}
\StepR{$=$}{Inductive hypothesis and $C\MM W= C'$}
{\nfold{N}{C'}\, \kron\JJ C'}
\Step{$=$}
{\nfold{N+1}{C'}~.}
\end{Reason}
\end{Proof}
\end{Lemma}
\NStuff{Immediate is that Kronecker product respects re\-fine\-ment-pres\-er\-va\-tion.}

\begin{corollary}\Label{c1053}
Let a set \CCC\ of channels be refinement-preserving. Then the set of its $N$-fold Kronecker products, written $\nfold{N}{\CCC}$, is also
refinement-preserving.
\end{corollary}



%


Further, we can use refinement-preservation of Kronecker product to deduce that random-response is also
refinement-preserving.

\begin{corollary}\Label{c1450X}
The set of $N$-participant and \KK-choice randomised response mechanisms $\nfold{N}{(\RRRR^\KK)}$ is
refinement-preserving.
\end{corollary}

\section{More general stable pipelines}
\Label{s1008}

\subsection{Proving stability in general}\Label{s1037}
Our aim now is to find (and prove correct) conditions on perturbation mechanisms and post-processors that will ensure stability when they are used together.

\NStuff{The following theorem gives sufficient
conditions for stability in terms of refinement-preserving families \CCC\ of perturbation mechanisms and certain sets \PPP\ of post-processors.}

\begin{Theorem}\Label{t1321}
\NStuff{
Given a refinement-preserving family of perturbation mechanisms   \CCC\  (described as channels), define \WWW\ to be the set of all stochastic ``refinement witnesses'' that can arise from refinement-pairs in \CCC. That is, for all $C, C'$ in \CCC\ with $C\,{\ChanRef}C'$ there is a $W$ in \WWW\ such that $C\MM W=C'$.\,}%

The two conditions below are then sufficient for stability of $(\CCC,\PPP)$, as given in \Def{d1632}:
%
for all $P$ in \PPP\ and $W$ in \WWW\ we have
\begin{enumerate}
\item\Label{i1321A} The post-processor $P$ has a left-multiplicative inverse \PI.\,%
\footnote{Note that $P$ does not have to be square to have (only) a left inverse.}
\item\Label{i1321B}
The post-processor $P$ and witness $W$ together satisfy
\begin{equation}\label{e1914}
	P{\MM}\PI\MM W\MM P \Wide{=} W\MM P \Q.
\end{equation}
\end{enumerate}~
\begin{Proof}
\NStuff{
\item\Label{i1519i} From $C,C'$ in \CCC\ and $C\,{\ChanRef}\,C'$ we have
that $C'\,{=}\,C\MM W$ for some $W$ in \WWW. Thus for stability (\Def{d1632}) we need only show that $C\MM P \ChanRef C\MM W\MM P$. For that, it suffices that $P\ChanRef W\MM P$ since it is clear
that left-multiplication by a channel preserves refinement (\Thm{t1516X}).} 

We now show that if $P$ is deterministic, has no all-zero columns, and has a left-inverse \PI\ (i.e.\ so that $\PI\MM P=\Id$\,), then
\begin{equation}\label{e1505}
	P\ChanRef W\MM P \WIDE{\textrm{if and only if}} P\MM\PI\MM W\MM P \;=\; W\MM P \Q.
\end{equation}
Although the right-hand side here looks more complex than the left, it is easier in practice to establish: for to establish a refinement between $P$ and $W\MM P$ (as on the left), one has to \emph{find} a witness. But on the right-hand side, we need only to check for equality.

We prove \Eqn{e1505} as follows:
\begin{description}
\item[if] This is trivial, because $\PI\MM W\MM P$ is the refinement witness.
(Recall \Lem{l1601X}.)

\item[only if]
If $P\ChanRef W\MM P$ then there is a witness $X$ such that $P\MM X = W\MM P$, and it is unique
(because $P$ has a left-inverse). For if we multiply both sides on the left by \PI, we get
\[
	\PI\MM P\MM X \Wide{=} X \Wide{=} \PI\MM W\MM P \Q.
\]
Substituting that $X$ into $P\MM X = W\MM P$ from above gives $P\MM \PI\MM W\MM P = W\MM P$,
as required at (\ref{e1914}).
\end{description}
\end{Proof}
\end{Theorem}

\NStuff{We remark that \Thm{t1321} can be seen as a generalisation of \Lem{l1601X} by noting that every random-response mechanism $R''$ is itself a refinement witness (for another random-response pair).}

Shortly we will use the {\bf if} in \Thm{t1321} for example to show that, in the general case $N{>}1$, the family \RRRR\ of random-response mechanisms and the set \TTT\ of counting post-processors are together stable. The {\bf only if} has a more subtle use, which is explained 
in \App{a1433X}:
with it, we can show (remarkably) that a \CCC/\PPP\ pair is definitely unstable in \emph{some} setting --- even without exhibiting the specific prior and loss function of the setting.

\subsection{The set \TTT\ of counting queries}\Label{s1350T}\Label{s1123}
We now define order-$N$ ``counting queries'' in matrix terms.
\begin{definition}\Label{d1517-z}
A \emph{coun\underline{T}ing query} $T$ is the $N$-Kroneckering of a ``Boolean aggregator'' $B$ followed by a ``\J tall\underline{Y}''$\;Y$. \Nf{I misread as "Tall Y". Can we name it something else?}

A \emph{Boolean aggregator} $B$ is a function (in matrix form) that takes every choice in \KK\ to either \true\ or \false.\,%
\footnote{We should write $B^\KK$, but we will assume $(-)^\KK$ throughout, to avoid clutter.}
A subsequent \emph{tally} $Y$ takes an $N$-tuple of \bool's to the number of \true's in it --- that is, it ``counts the \true's''. Thus a counting query overall, some $T= \nfold{N}{B}\MM Y$, takes an $N$-tuple of \KK's as input, and outputs how many elements of the tuple would be mapped to \true\ if acted on by $B$ individually.
\end{definition}
The types above are $B$ of type \Matrix{\KK}{\bool}, and thus $\nfold{N}{B}$ of type $\Matrix{\KK^N}{\bool^N}$, and $Y$ of type $\Matrix{\bool^N}{\Nat}$ and thus the whole counting query $T=\nfold{N}{B}\MM Y$ of type $\Matrix{\KK^N}{\Nat}$.
An explicit definition of $Y$ is
\[
	Y_{x,n} \Wide{=} \textrm{$1$\quad if the number of \true's in tuple $x$ is $n$ else\quad0}\Q.
\]

With the above preparation, we can now look at the more general case of stability for $(\CCC,\PPP)$ comprising $\CCC{=}\RRRR$ random-response perturbers and $\PPP{=}\TTT$ counting post-processors.
\subsection{Extension to $N$-way perturbers }\Label{s1347X}

Our aim is to show how to apply  \Thm{t1321} to the more challenging scenarios where there are many participants contributing perturbed data. The next lemma demonstrates how to do this: it is a Kroneckered $N$-way version of our technical \Lem{l1601X} and a restriction from \PPP\ to \TTT.
\begin{Lemma}\Label{t1201X}
If \J\J\J$T$ in \TTT\ is a counting query (\Def{d1517-z}) and $\nfold{N}{W}$ is an $N$-way Kroneckered witness for refinement between two $N$-Kroneckered random responses over \KK, i.e.\ a Kroneckered random response itself (\Cor{c1450X}), then
\begin{equation}\label{e*1156}
	T{\MM}\TI\MM \nfold{N}{W}\!\MM T \Wide{=} \nfold{N}{W}\!\MM T \Q.
\end{equation}

Parentheses are used (in spite of associativity), or extra space, to indicate in each step the portion of the expression currently of interest: we reason
\Cf{The spaces and parentheses might not be quite right.}
\begin{Reason}
\Step{}{T{\MM}\TI\MM \nfold{N}{W}\!\MM T}
\StepR{$=$}{\Def{d1517-z} of $T$}{(\nfold{N}{B}\!\MM Y){\MM}\BBI\MM \nfold{N}{W}\MM (\nfold{N}{B}\!\MM Y)}
\WideStepR{$=$}{Left inverse of product}{\nfold{N}{B}\MM Y{\MM}\quad(\YI\MM\KKR)\quad\MM \nfold{N}{W}\MM \nfold{N}{B}\MM Y}
\WideStepR{$=$}{Left inverse of Kronecker}{\nfold{N}{B}\MM Y{\MM}\YI\MM\nfold{N}{(\BB)}\MM \nfold{N}{W}\MM \nfold{N}{B}\MM Y}
\WideStepR{$=$}{Distribution of Kronecker}{\nfold{N}{B}\MM\quad Y{\MM}\YI\MM\nfold{N}{(\BB \MM W \MM B)}\MM Y}
%
%
\StepR{$=$}{See below (\J$\dagger$)}{ \nfold{N}{B}\MM\quad\nfold{N}{(\BB \MM W \MM B)}\MM Y }
\StepR{$=$}{Distribution of Kronecker}{ \nfold{N}{(\quad B\MM\BB\MM W \MM B\quad)}\quad\MM Y }
\StepR{$=$}{See below (\J$\ddag$)}{ \nfold{N}{( W \MM B)}\MM Y }
\StepR{$=$}{Distribution of Kronecker}{\nfold{N}{W}\MM \nfold{N}{B}\MM Y }
\StepR{$=$}{Definition of $Y$}{W\MM Y~.}
\end{Reason}

\begin{enumerate}
\item[($\dagger$)] To show here is that
\[
	Y{\MM}\YI\MM\nfold{N}{(\BB \MM W \MM B)}\MM Y
	\Wide{=}
	\nfold{N}{(\BB \MM W \MM B)}\MM Y\Q,
\]
and it is proved in \Lem{l1226X} further below, where we use the fact that
$\BB \MM \RR \MM B$ is of the very simple type $\Matrix{\bool}{\bool}$ and moreover is stochastic, i.e.\ it is a $2\times 2$ channel.
\item[($\ddag$)] To show here is that
\begin{equation}\label{e1556}
	B\MM\BB \MM W \MM B \Wide{=} W \MM B \Q,
\end{equation}
whose form indeed looks very like our original goal \Eqn{e*1156}. But it is not circular reasoning, because \emph{here} we have the Boolean aggregator $B$ (rather than the more general $T$), whose output type is \bool; and $W$ is not Kroneckered. Thus it is the simple case that we have already proved (\Lem{l1601X} of \Sec{s1347X}).\,%
\footnote{The reason \Eqn{e*1156} does not follow directly from Kroneckering \Eqn{e1556} is that the $T$ of \Eqn{e*1156} is not the $\nfold{N}{(-)}$ Kroneckering of anything: it contains the tally $Y$ as its second component, and $Y$'s not being a Kroneckered matrix is precisely its purpose --- it is \emph{not} applied to each tuple-component separately: rather it depends on the tuple as a whole.}%
\end{enumerate}
\end{Lemma}
An interesting further corollary of \Lem{l1601X} however is that it can be used as a technique to prove positive results for perturbers \emph{other than} random response. In particular, provided the batching aggregator $B$ inside the counting query $T$ preserves stability for a single perturber instance, then stability is preserved also for a corresponding Kroneckered perturber. \Sec{ss1634} gives an example.

\begin{lemma}\Label{l1226X} (used to prove ($\J\dagger$) above)
Let $C$ be a $2{\times}2$ stochastic matrix (i.e.\ a channel) and $Y$ a tally as defined above. Then we have
\begin{equation}\label{e1016-a}
Y{\MM}\YI\MM \nfold{N}{C}\MM Y \Wide{=}  \nfold{N}{C}\MM Y\Q.
\end{equation}
\begin{proof}
The full details of the proof appear in \App{a1154}.
\end{proof}
\end{lemma}

\section{Stability and instability in Random Response perturbations}

\subsection{Stability for $N$-way \RRRR\ and counting \TTT}\Label{s1127}
We now use the results of the previous section to prove our first general positive result: that even in the many-respondent case, i.e.\ with $1{<}N$, when a pipeline combines a random response mechanism
with a post-processing counting-aggregator, then it is stable:

\begin{theorem}\Label{t0923}
The pair $(\nfold{N}{\RRRR},\TTT)$ is stable.
%
\begin{proof}
We show that all conditions for \Thm{t1321} apply. From \Cor{c1450X} we have that  \nfold{N}{\RRRR}  is refinement-preserving; next from \Def{d1517-z} that $T{\in} \TTT$ is deterministic, and therefore has a left inverse. Finally \Lem{t1201X}  shows that the equality \Eqn{e1914} of \Thm{t1321} holds for any $\nfold{N}{R}\in \nfold{N}{\RRRR}$ and $T\,{\in}\,\TTT$.
%
\end{proof}
\end{theorem}

The above says that for a pipeline $\nfold{N}{\RR}\!\MM T$, i.e.\ one using random response for perturbation followed by a counting query, decreasing the \EDP\ satisfied by its perturber (increasing privacy) always results in a decrease in utility, for \emph{any} utility modelled by a loss function as defined in \Sec{s1150X}, and any prior on \XX.
\subsection{\underline{In}stability for $N$-way $\RRRR$ and \underline{summin}g $S$}
\Label{s0915X}

We now give a negative result: a mechanism formed from a random-response perturber followed by a \emph{sum} aggregator is not necessarily stable.

Consider the following random-responses defined for choice $\KK{=}\{0,1,2\}$, where we write $R_2$ for $\RR[\,\ln2]$ and $R_3$ for $ \RR[\,\ln3]$ :
\[\small
     R_2 =
     		\begin{bmatrix}
               1/2 & 1/4 & 1/4 \\
               1/4 & 1/2 & 1/4 \\
               1/4 & 1/4 & 1/2
            \end{bmatrix}~~\quad\quad
     R_3 =
     		\begin{bmatrix}
             3/5 & 1/5 & 1/5 \\
             1/5 & 3/5 & 1/5 \\
             1/5 & 1/5 & 3/5
           \end{bmatrix}
           \Q.
\]

Observe that $R_2$ realises \EDP\ of $\ln 2$ exactly and $R_3$ realises \EDP\ of $\ln 3$, so that $R_3 \ChanRef R_2$ because they are both 
in the same random response family \cite{Chatzi:2019}.
\Cf{We could also show this directly, using a witness: it would also illustrate that the witness is indeed RR itself. Looks like $W$'s truth-telling probability $p$ would be \NF{5}{8}, giving
\[
     W \Wide{=} \begin{bmatrix}
                   6/8 & 1/8 & 1/8 \\
                   1/8 & 6/8 & 1/8 \\
                   1/8 & 1/8 & 6/8
                \end{bmatrix}\Q,
\]
and $S\MM W = R$.
Wouldn't that be a nice example? \GStuff{Could we keep both?}}

Now assume there are two data holders sharing data, so that the perturber is a $\kron2$-Kronecker product: it is either  $\nfold{2}{R_2}$ or $\nfold{2}{R_3}$. The following defines a post-processing \emph{sum} aggregator $S$:\,%
\footnote{There are many counting post-processors in \TTT, depending on the aggregator used; but here there is only one summing aggregator $S$.}
\begin{equation}\label{e1526X}
S_{v, n} \Wide{\Defs} 1~~\textit{if}~~ \textit{sum}(v)=n~~\textit{otherwise}~~0~,
\end{equation}
%
where $S$ has rows labelled by  the (perturbed) outputs from $R_2$ or $R_3$ --- notice that they form a pair representing the complete output from the Kroneckered product that become inputs for the post-processing summing analyst $S$. The column labels are the analyst's statistics:  in this case she is interested in computing the sums for each input vector, so $S_{v, n}$ is $1$ exactly when the sum of $v$'s entries is equal to $n$. Thus $\textit{sum}(1,0)$ returns $1{+}0=1$, and $\textit{sum}(1,2 )$ returns $1{+}2=3$.
For this $P$ we find that $\nfold{2}{R_3}\!\MM S\;{\ChanRefN}\;\nfold{2}{R_2}\!\MM S$. 
(See \App{a1531} for details.)
\,%
\footnote{The refuting loss function can be obtained using the {\tt libqif} library from \url{https://github.com/chatziko/libqif}. 
}

What this means is that we can find a sum-querying analyst who prefers random-response obfuscation $R_2$ to $R_3$ for a sum-query release, \emph{even though the $\VE$ for $R_2$ is lower (more private) than for $R_3$}.

{
\section{Stability and instability in geometric perturbations}\Label{s1215X}
\Af{I would add in another positive case here: will do it at the end. See \Sec{s0936}.}

We now consider the behaviour of counting queries as post-proc\-es\-sors for \emph{geometric} mechanisms $G$ --- that is, pipelines $M=G\MM P$ where $G$ is a geometric mechanism and $P$ is either a counting query $N$ or a summing query $S$. Some settings in which they arise are in geo-location privacy~\cite{andres2013geo, DBLP:journals/corr/abs-1906-12147}, and in machine learning via the Noisy \ArgMax\ mechanism, which we encounter in \Sec{sec:argmaxX}.


%
It has been shown
that the \emph{truncated} geometric mechanism
(\Def{d2351X} above)
and the (infinite) geometric mechanism are equivalent under refinement \cite{Chatzi:2019}. Because of that, they contribute equivalently to any of the pipelines we study here.

\subsection{In/stabilities in counting and summing}\label{ss1634}
\Label{s0952X}


In contrast to random-response perturbation, geometric perturbation is \emph{not} in general stable wrt.\ either counting or sum queries.  We present the following counter-examples.

The two geometric mechanisms $G_2 (= \GD[\,ln 2])$ and $G_3 (= \GD[\,ln 3]) $ on the domain $\{0,1,2\}$ are defined
\[\small
     G_2 =  
     	   \begin{bmatrix}
               2/3 & 1/6 & 1/6 \\
               1/3 & 1/3 & 1/3 \\
               1/6 & 1/6 & 2/3
            \end{bmatrix} ~~\quad\quad
     G_3 = 
            \begin{bmatrix}
             	3/4 & 1/6 & 1/12 \\
             	1/4 & 1/2 & 1/4 \\
             	1/12 & 1/6 & 3/4
           \end{bmatrix}\Q,
\]
where the ``$\alpha$'''s used in \Def{d2351X} to achieve the \VE's shown are \NF{1}{2} and \NF{1}{3} repectively.
As for random responses, since $G_{2, 3}$ are in the same family of geometric perturbers we have that $G_3\CR G_2$ because $\ln2{<}\ln3$ \cite{Chatzi:2019}.\,%
\footnote{Here the \VE's are computed wrt.\ the Euclidean metric on the domain; but in fact for \emph{every} metric (including the discrete metric), the same ordering holds. That comes from the stronger refinement property, as shown in \cite{Chatzi:2019}.}

Consider \emph{two} participants, each applying geometric perturbations independently to their own data,
and a subsequent counting query $N$ that counts the number of $0$'s in the output \YY.  The noisy geometric perturbation in this case is a Kronecker product, either $\nfold{2}{G_2}$ or $\nfold{2}{G_3}$.
Now suppose that the analyst wishes to count the total number of $0$ results. Recall from \Def{d1517-z} that this is equivalent to defining an aggregator over the output of the perturbed input and then tallying the number of true results. Here we use aggregator $Z$:
\[
Z_{y, t}  \Wide{\Defs} ~~ [y=0]~,~~\textit{and}~~~~\quad Z_{y, f}  \Wide{\Defs} ~~ [y\neq 0]~,
\]
where we use Iverson brackets to map true to $1$ and false to $0$. With this definition the corresponding pipelines are \emph{not} in refinement: though $\nfold{2}{G_3}\ChanRef \nfold{2}{G_2}$, we have $\nfold{2}{G_3}\MM (\nfold{N}{Z}\MM Y) \ChanRefN \nfold{2}{G_2}\MM  (\nfold{N}{Z}\MM Y)$, where $Y$ is the tally operation from \Def{d1517-z}.
%
%
%



On the other hand, if the analyst only wishes to count outliers \emph{of either kind}, she would use an aggregator as in \Def{s1123}, i.e.
\[
L_{y, t}  \Wide{\Defs} ~~ [y\in \{0, 2\}]~,~~\textit{and}~~~~\quad L_{y, f}  \Wide{\Defs} ~~ [y\not\in \{0, 2\}] \Q,
\]
where the batching groups the noisy outputs into two classes: an ``outlier'' with value $0$ or $2$, or a ``non-outlier'' with value $1$.   In this case, it turns out that $L$ and $\nfold{N}{L}$ satisfy the conditions for \Thm{t1321} wrt.\ the witness for refinement for $G_3 \ChanRef G_2$ --- and thus we do have that $\nfold{N}{G_3}\MM (\nfold{N}{L}\MM Y) \ChanRef \nfold{N}{G_2}\MM  (\nfold{N}{L}\MM Y)$. 
(See \App{AS1553}.)

\vspace{0.5cm}

Similarly, we can define a sum-query post-processor to be the $S$ at \Eqn{e1526X},
finding this time that
$\nfold{2}{G_3}\MM S \ChanRefN \nfold{2}{G_2}\MM S$\,.
%
This means that, for some decision-theoretic analysts (modelled using an appropriate gain/loss function and prior), the utility they get from data releases using channel $G$ as the randomising mechanism is better than the utility obtainable from channel $H$, even though $G$ is more private than $H$ (that is $\VE_G{<}\VE_H$). In this case the analyst and the data holders both win --- more privacy yields more utility. However, the above does not tell us for \emph{which} analysts this behaviour occurs. In \App{s1102X}
it is shown experimentally that for a well-known loss function (the mean error) the unstable behaviour actually does occur: that is, we find that \emph{increasing} privacy can, for some values of $\VE$, \emph{also increase utility}.
} 

\section{The instability of ``noisy \ArgMax''}\Label{sec:argmaxX}

\subsection{The structure and purpose of noisy \ArgMax}\Label{a1313}
\Alg{a1004X} below is a standard implementation of a privacy-preserving program that, given a histogram (created from some dataset), calculates (internally) a perturbation of that histogram and outputs the mode of the resulting perturbed histogram. We can model its information-flow characteristics with a pipeline as follows.

That noisy \ArgMax\ pipeline has \emph{three} stages (unlike those we have studied so far, which had only two). There is as usual a perturber (geometric) and an \ArgMax\ post-processor $A$, but now as well there is a pre-processing stage which we will call $H$ (for ``histogram''). Thus for noisy \ArgMax\ the full pipeline is $M=H\MM G\MM A$.

\emph{The first stage} $H$ takes as input an $N$-tuple of choices from a set \KK\ of size $K$  (as usual for our pipelines); thus the size of the whole pipeline's input space is $K^N$. Each of those $N$-tuples is converted into a histogram giving, for each possible choice in \KK, the number of participants that made that choice, i.e.\ for each $k$ in \KK\ some number $0{\leq}n{\leq}N$. And each tuple sums to the original $N$. That is the space \XX\ that is passed from the pre-processor $H$ to the geometric perturber $G$.

The reason we do not combine $H$ and $G$ into a single perturber $H\MM G$ is that our utility (when we consider it) will \emph{not} refer to the original $N$-tuple of choices given to $H$: instead it will refer to the histogram in \XX\ that passed from $H$ to $G$ --- and so that cannot be ``hidden'' within a composition $H\MM G$.



\emph{The second stage} $G$ takes one of those histograms $x$, 
and applies the geometric perturbation independently to each of its bars, to produce a perturbed output histogram. Just as above, we can express that as a Kronecker product of solitary perturbations --- but note that it is now a $K$-fold Kronecker product (i.e.\ not $N$-fold as in earlier examples).
Remarkably, the use of Kronecker here summarises the complete information-flow channel's acting on a whole histogram --- but each one of them is a possible input histogram for \Alg{a1004X}, thus producing a row of the channel $\nfold{K}{(\GD)}$. The columns of the channel $\nfold{K}{(\GD)}$ correspond to the possible perturbed output histograms.

\emph{The third stage} $A$ takes a perturbed output histogram $y$ from $G= \nfold{K}{(\GD)}$, and applies the \ArgMax\ algorithm to it, which as usual can be expressed as a matrix $A$,\,%
determining a value $k$ in \KK\ that was maximally chosen among the $N$ original participants.\,%
That is, it finds a $k$-bar whose height is the \emph{mode} of the original values. Thus the rows of $A$ correspond to a complete histogram (as input) and the columns correspond to the mode.\,%
\footnote{Although the mode is unique, there might be several $k$'s that realise it. \ArgMax\ gives one of them.}
Any output from \Alg{a1004X} corresponds to the output of the pipeline model.

And the \EDP\ satisfied by  the  whole pipeline is immediate:
\begin{lemma}\Label{l1532}
The mechanism $\;\nfold{K}{(\GD)}\MM\ArgMax\;$ as a whole is \EDP\ for the same \VE\ used in its construction.
\begin{proof} $\nfold{K}{\GD}$ satisfies \EDP, and post-processing by \ArgMax\ (indeed by anything) preserves that, by the \DPI.
See \App{AS1623} for details.
\end{proof}
\end{lemma}

\begin{algorithm}
\begin{verbatim}
# Assume here that the choice type is k in 0<=k<K,
# with H[0:K] therefore our input of type X:
# a histogram produced by a pre-processing step.
# The value of K is known but the bar-heights,
# for each 0<=k<K, are not.

# Apply (truncated) geometric noise to each bar.
# Parameter eps can be varied to select an
# appropriate level of privacy.
for k in range(K): H[k]= H[k]+GD(eps,K)

# Locate position of maximum value in (non-empty) H,
modePos= 0         # i.e. implement standard ArgMax.
for k in range(1,K):
    modePos= (k if H[k]>H[modePos] else modePos)
print(modePos)
\end{verbatim}

\bigskip
\flushleft
Note that although the number of participants in the original data is $N$, it is the number of choices $K$ in the original data that has become the number of ``participants'' after \verb'H''s pre-processing: each choice $k$ is a participant; and each bar-height \verb'H[k]', between 0 and $N$ inclusive, is its choice.

\caption{Noisy \ArgMax\ using the geometric mechanism}\label{a1004X}
\end{algorithm}

In summary: \Alg{a1004X} assumes the conversion to a histogram is found in  \verb'H[0:K]' (i.e.\ that the pre-processing $H$ has already been done). That histogram's bars are independently geometrically perturbed in the first part of the algorithm (by \GDX); and then in the second part the \KK-label of a tallest bar of the perturbed histogram is output.  Because in certain scenarios delivering the precise position of the \emph{actual} mode (i.e.\ of the original histogram) has some privacy issues~\cite{papernot2016semi}
a differential-privacy style perturbation is applied as indicated here.
\footnote{Taken from {\tt https://programming-dp.com/notebooks/ch9.html}~.}
Outputting the label of the tallest bar is in effect estimating the \emph{mode} of the values in the pipeline's input $N$-tuple of \KK's.

\subsection{Evaluating the utility of noisy \ArgMax}
The utility of noisy \ArgMax\ depends on the difference between the mode of its input histogram and the mode of the histogram it outputs after having applied (geometric) perturbation: as usual, that utility is expressed as a loss function, in this case having as its two arguments the input histogram and the output mode.


Noisy \ArgMax\ implemented with geometric perturbation, as in \Alg{a1004X}, is unfortunately not stable in general: that means that in some circumstances increasing the {\tt eps} parameter in \Alg{a1004X} might \emph{not} increase the overall utility, although (\Lem{l1532}) it will always reduce the privacy. The following makes this precise:
\begin{lemma}[Instability in noisy \ArgMax]\Label{l1704}

Even if $\VE{>}\VE'$ (so that $\GD\CR\GD[\VE']$, suggesting decreased utility), we find that the \GD[\relax]-then-\ArgMax\ pipeline-refinement fails:
\[
\nfold{K}{(\GD)}\MM\ArgMax \Wide{\ChanRefN} \nfold{K}{(GD_{\VE'})}\MM\ArgMax ~.
\]
\begin{proof}
We use the the {\bf only if} referred to in the proof of \Thm{t1321}, in combination with the special properties of witnesses for geometric-mechanism refinement (\Sec{s0914}). 
See \App{AS1623} for details.
\end{proof}
\end{lemma}

\section{Stability experiments}\Label{a1454X}
\subsection{Noisy \ArgMax: a closer look}\Label{s1421}

In this section we show that, in spite of \Lem{l1704}, there are important loss functions $\ell$ that measure the utility of interest specifically for noisy \ArgMax, and further that experiments imply that they seem to be $\ell$-stable. The loss function \AMA\
defined next is one such example. It measures the ability of the analyst to determine the correct modal value of the input histogram.


\begin{definition}
Let $N$ be the number of participants, and let each participant make a choice from the same set \KK\ . After pre-processing makes those choices into a histogram, our input set \XX\ to be perturbed will be of type $\KK\Fun(0..N)$. The output set \YY\ is, as usual, of the same type: again a histogram, but not quite the one that was input. The post-processing step then converts such a $y$ in \YY\ into a choice $k'$ that is supposed to be (close to) the mode $k$ of the original data (i.e.\ the \ArgMax\ of the original $x$ in \XX\,).

We define our loss function \AMA\ for assessing the accuracy of \ArgMax\ (``\ArgMax\ Accuracy''), as follows:
\begin{equation}\label{e1717}
	\AMA(x,k') \WIDE{=} \textrm{0\quad if~$\ArgMax(x)\,{=}\,k'$~else\quad 1} \Q.
\end{equation}
\end{definition}
The analyst loses 0 if she guesses the mode correctly, but loses 1 if she guesses wrong.
\Cf{I don't yet understand how \LF\ deals with multiple mode indices. Intriguingly, the \ArgMax\ function takes $x$ as its first argument (not $k$), and so could apply a more sophisticated evaluation of ``how good'' an answer $k'$ is, say ``How close is $h(k')$ to the actual mode of $x$?''
}

\begin{figure}[!ht]
		\centering
		\includegraphics[width=0.3\textwidth]{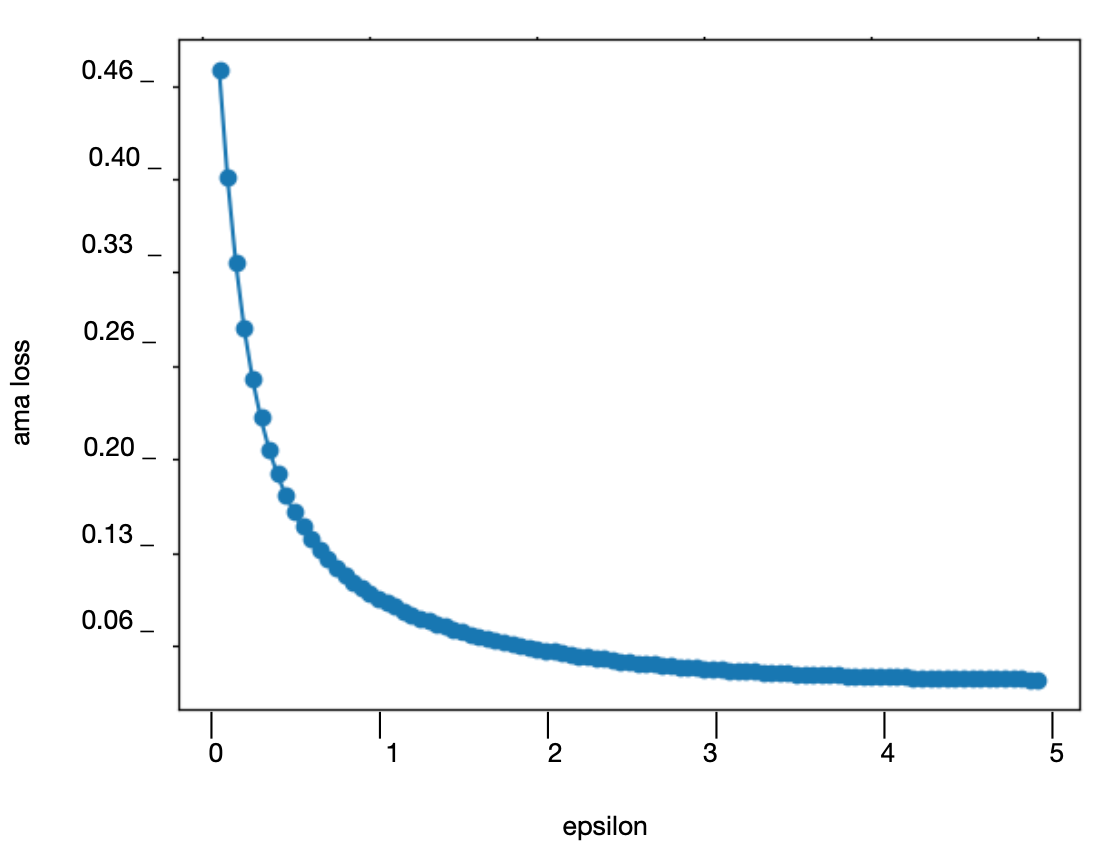}

		Noisy \ArgMax\  \CStuff{appears to be}  \AMA-stable; \AMA\ is given at \Eqn{e1717}.
	\caption{{Noisy \ArgMax\ for 20 individuals, 3 categories.}}\label{fig:argmax_expX}
\end{figure}

Consider a scenario where 20 participants are given the names of three celebrities and each is asked to pick their favourite one.  In order to preserve privacy the histogram of choices is subjected to \Alg{a1004X}; the analyst is then required to determine which celebrity received the most votes. In this scenario, the Bayesian analyst would use the loss function \AMA\ to guide her choice since, for any output from \Alg{a1004X} \AMA\ indicates the most likely mode (of the input histogram) that corresponds to the received observation.
Fig.~\ref{fig:argmax_expX} shows the change in average \AMA-utility, i.e.\ $U_\AMA(\pi, \nfold{3}{(\GD)}\MM\ArgMax)$ with the variation of the $\VE$-parameter in \Alg{a1004X}. Here we are assuming that the analyst does not know the disposition of the participants, and so she uses a uniform distribution over the $105=(21{\times} 20/2)$ possible histograms.

What we observe is that, for this particular loss function, \Alg{a1004X} \CStuff{appears to be} an \AMA-stable mechanism: the graph depicted shows that the utility improves (i.e. \AMA-loss decreases) as the privacy is reduced (increasing \VE), and in a way that can be relied upon. In an experimental setting this means that the analyst can locate relatively easily the optimal setting of $\VE$ to give an acceptable accuracy for her specific application of interest. In view of \Lem{l1704}, if she chose a different loss function to focus on a different aspect of the histogram, for example if she wanted to minimise the ``distance'' from the true mode (in the case that the categories are numeric), she might not be so fortunate.

\Cf{\Gz\Nz The \Fig{fig:argmax_expX} $y$-axes go the opposite way to Gabriel's in \Fig{fig:experimental-resultsA} and \Fig{fig:experimental-resultsX}. Should we invert the graphs and Label the $y$ as ``Bayes Risk''? \AStuff{yes, have done that and used the terminology we set up earlier.}}

\subsection{Instability ``in the wild'': the public domain}\label{ss1554}
\Label{s0959X}

\begin{figure}[!bt]
	\centering
	\vspace{-08mm}
	\begin{subfigure}[h]{0.45\textwidth}
		\centering
		\includegraphics[width=\textwidth]{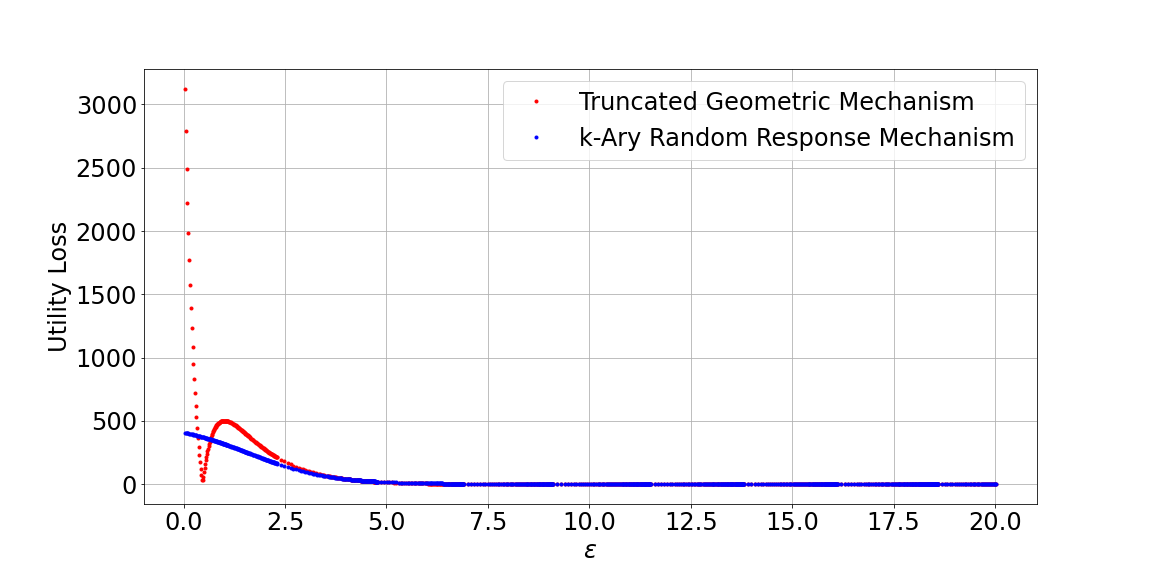}
		\caption{Complete results for $0.0 < \VE \leq 20.0$.}
		\label{fig:experimental-results-full}
	\end{subfigure}
	\begin{subfigure}[h]{0.45\textwidth}
		\centering
		\includegraphics[width=\textwidth]{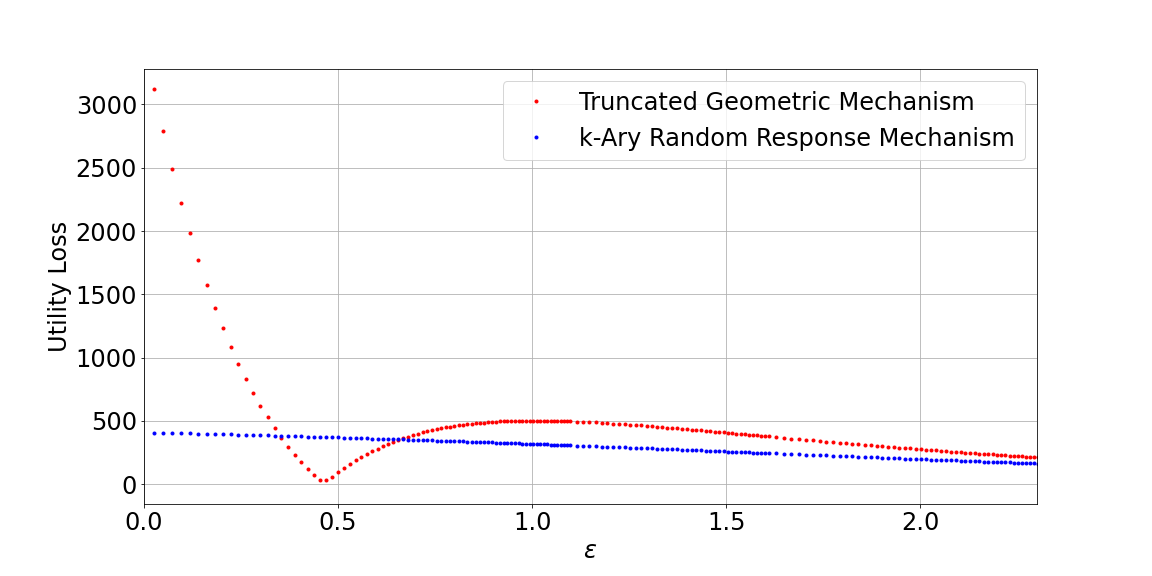}
		\caption{Partial results for $0.0 < \VE \leq 2.3$.}
		\label{fig:experimental-results-partial-0.0-2.3}
	\end{subfigure}

	\caption{Example of instability in the public domain}
	\label{fig:experimental-resultsA}
\end{figure}
Our next example is taken from the public domain, an investigation of the relationship between
utility and privacy in the general scenario of Fig.~\ref{fig:post-processing-pipelineX}.
\Af{, and reported elsewhere \cite{DBLP:conf/concur/AlvimFMN20} we cannot refer to this because it is against the anonymous reviewing rules}
We used a publicly available dataset released by ProPublica,\,%
\footnote{\href{https://github.com/propublica/compas-analysis}{\url{https://github.com/propublica/compas-analysis}}}
which contains two years' worth of data from the \COMPAS\ tool,\,%
\footnote{\COMPAS\ is a commercial assessment tool for the evaluation of re-offense by criminal defendants. The tool is popular in the United States criminal-justice system, but was shown by ProPublica to have significant biases towards black defendants.}
to produce the graphs of \Fig{fig:experimental-resultsA}.

The data were pre-processed for consistency,
resulting in a dataset with 11,710 unique records.\,%
\footnote{All but the most recent record for any given person were eliminated, as well as all records with invalid entries for
``marital status'' and ``risk of failure to appear''.}
Each individual record had its ``custody status'' attribute converted to a number
and then perturbed by the
truncated geometric mechanism \GD\ (Def.~\ref{d2351X}) of differential privacy.\,%
\footnote{This is reasonable because the numbering was chosen in proportion to the ``seriousness'' of each possible value for ``custody status'':
	0 for ``pretrial defendant'',
	1 for ``residential program'',
	2 for ``probation'',
	3 for ``parole'',
	4 for ``jail inmate'', and
	5 for ``prison inmate''.
	Indeed, it can be shown that this is equivalent to using an exponential mechanism with a well suited loss function; but the details are not relevant here.}
The collection of perturbed data was then shuffled, so the link between each data subject and their randomised value was lost.
The counting query \emph{\qm{\J Count the number of individuals with `custody status' value equal to `pretrial defendant'\J}} was then
performed as a post-processing of the resulting collection of
perturbed and shuffled values.
%
Consider that a data analyst seeing the result of
this query has as prior knowledge the original values for all
individuals' ``custody status'' \emph{except} for a single individual
considered to be the target.
The analyst assumes, also as prior knowledge, that this individual's
unknown value has the same distribution derived from the frequency of the
known values, and that her goal is to find out this individual's value.
Her chosen utility measure is the posterior Bayes' risk (Eqn.~\Eqn{BR}) of the counting query on the real data given the reported value for that same query on the perturbed data.

\Fig{fig:experimental-resultsA} presents the results of
the experiment in the above scenario for various values
\Mf{We need to clarify here that QONFEST did not report instability. In that work we skipped the weird values. We need to say that we use QONFEST's code ourselves and we found the instability we report here. Does that check, \GStuff{Gabriel} and \NStuff{Natasha}?}%
\Gf{\MStuff{Mario}, instability was not reported there, but I think this would not make sense because the QONFEST code is not available anywhere yet.}%
\Nf{Check footnote above.}%
of \VE; and we can see from those results that this combination --the \GDX\ perturbation and subsequent counting--
is unstable in the exact sense defined in 
the main body of this paper:
it is \emph{not order-preserving} in expected loss ($y$ axis) as a function of the \DP\ parameter \VE\ ($x$ axis).
\Mf{\Gz, we need to make these graphs color-blind and printer-friendly. We need to have each curve to have its own symbol (e.g., +, o, x, triangle, etc), and to be visible in gray-scale. Also, why do we have the blue line corresponding to K-RR?
\Nz, do we want to make a point that k-rr is stable? If so, we should say it explicitly and link back to the section that says that.}
More precisely, as \VE\ grows, the utility loss of the geometric mechanism decreases and then increases again, as depicted in Fig.~\ref{fig:experimental-results-full}, and
in more detail in Fig.~\ref{fig:experimental-results-partial-0.0-2.3}.
\MStuff{For the sake of comparison, we also present the results for an experiment in which the $N$-way random response of
\Sec{s1127} is applied locally instead of the truncated geometric mechanism \GD:
as predicted by \Thm{t0923}, this scenario is stable and utility is monotonic wrt.\ privacy.}

In our experiments we originally regarded this as an anomaly whose explanation could have been either a bug in our analysis code, or a consequence of rounding errors in the arithmetic. After investigating these possibilities, and finding that neither seemed to be the cause,  we sought a concrete way to confirm that \Fig{fig:experimental-resultsA} was indeed the correct representation of the actual behaviour of the scenario. In order to do that we created  a minimal counterexample for this scenario, using exact arithmetic, that accounts theoretically for the unstable behaviour we observed. 


\section{Discussion}\Label{s1120}
The literature on formal proofs for privacy is very broad, with some techniques applicable to actual implementations in code \cite{DBLP:conf/aplas/McIverM19,DBLP:conf/csfw/BartheGAHKS14,Barthe:16}. Less attention has been paid however to utility, and in particular to its fine-grained behaviour in relation to $\VE$. The application of refinement has been found to be useful for studies of optimality \cite{DBLP:conf/csfw/FernandesMPD22} in the sense of Ghosh et al.\ \cite{Ghosh2009}.  Here we found that the notion of stability appears to be another useful property indicating that complex mechanisms are behaving well from the point of view of an analyst. This is because the presence of unstable behaviour means that the post-processing part of the mechanism is actually losing information that is useful to the analyst: when there is no post-processing at all there is never any unstable behaviour.\,%
\footnote{This comes from \Def{d1526X}.}
 (That means the post-processing has incorrectly collapsed columns of the perturbing channel.) In particular, instability could indicate bugs in the post-processing algorithm, and so testing for it could be used as a useful debugging tool, complementing analysis of privacy \cite{Ding_2018}.

Further, any fine-grained analysis must be tractable: what we have shown here is that the Kronecker product and its algebra enables precise modelling of Bayesian properties and simple algebraic-style proofs --- from which we can draw strong and very general conclusions about utility, as well as simplifying proofs that verify privacy. The Kronecker product has also been used in modelling complex networks and data \cite{10.5555/1756006.1756039,10.1214/22-SS139} and we believe its further development for understanding privacy and utility will be similarly fruitful.

Finally, the assumption that \VE\ controls the trade-off between privacy and utility in \EDP\ is widespread, and often openly endorsed by seasoned researchers and members of governmental statistical agencies. For example, Ghandi and Jayanti  remark: ``If the privacy loss parameter [in \EDP] is set to favor utility, the privacy benefits are lowered (less “noise” is injected into the system); if the privacy loss parameter is set to favor heavy privacy, the accuracy and utility of the dataset are lowered (more “noise” is injected into the system)'' \cite[p.10]{GA2020}. Our concept of stability is, to the best of our knowledge, the first rigorous formalization of this assumption, and we were able to identify situations in which it does hold, and does not hold. We believe this is a relevant contribution to various privacy practitioners and policy-makers that design and evaluate ever more complex private systems, and that it may allow for the construction of simultaneously more private and useful pipelines. As some of our results suggest, a possible cost of stability is that some \DP-mechanisms may not be compatible with some post-processing algorithms. However, we do not yet know whether, under a different encoding of the privacy pipeline, stability could be in principle recovered. This question is left as future work.

\section{Related work}

\Af{We can mention Kuifje as being able to interpret exactly the code given in the figures. \GStuff{Done at the end of the section.}}




Differential privacy was first proposed by Dwork et al. in 2006 \cite{Dwork2006} and presents closure for privacy loss under post-processing \cite{Dwork2013}. The local model was first formalized by Kasiviswanathan et al. in 2011 \cite{Kasiviswanathan2011} as a generalization of randomised response, first proposed by Warner in 1965 \cite{Warner1965} and motivated by scenarios in which respondents do not trust the data curator.

Motivated by the reconstruction of statistical datasets via sequential queries proposed by Dinur and Nissim in 2003 \cite{Dinur2003}, the United States Census Bureau (USCB) started the adoption of differential privacy in 2008 \cite{Machanavajjhala2008} and developed the TopDown algorithm for the 2020 Census \cite{Abowd2018,Abowd2022}. The final version of the TopDown algorithm was implemented using the discrete Gaussian mechanism \cite{Canonne2020}, but previous demonstration data products based on data from the 2010 Census used the geometric mechanism \cite{Ghosh2009} instead. The choice of the discrete Gaussian over the geometric mechanism was the result of empirical accuracy tests of the TopDown algorithm using each mechanism with comparable privacy-loss budgets \cite{Abowd2022}. The budget allocation for the geometric mechanism under pure differential privacy (i.e. $\delta = 0$) was determined using $\VE$-allocation for a total of $\VE = 6$ \cite{Abowd2022}, while for the discrete Gaussian mechanism under zero-Concentrated Differential Privacy (zCDP) it was determined using $\rho$-allocation, with equivalent $\VE$-allocation ranging from $\VE = 4.36$ to $\VE = 19.61$ (the production value \cite{UnitedStatesCensusBureau2021}).

The choice of values for the parameter $\VE$ has been discussed since the inception of differential privacy and is usually regarded as a \qm{social question} \cite{Dwork2008}, with values of $\VE$ often ranging from $0.01$ to $10$ but without solid grounds for those choices \cite{Hsu2014}. One of the main challenges concerns the impact of the introduced noise, whose original goal is to protect respondents' privacy, on data accuracy and its usefulness to data analysts. This notion of an inverse relationship, or a trade-off, between privacy preservation and accuracy has been widely reported on the literature, e.g.: Dwork and Roth state in \cite{Dwork2013} that \qm{A smaller $\VE$ will yield better privacy (and less accurate responses)}, Hsu et al. state in \cite{Hsu2014} that \qm{$\VE$ is a knob that trades off between privacy and utility}, and a report commissioned by the USCB \cite{JASON2020} state that \qm{The trade-off between confidentiality and statistical accuracy is reflected in the choice of the DP privacy-loss parameter. A low value increases the level of injected noise (and thus also confidentiality) but degrades statistical calculations}.

Data-release pipelines such as the one in \Fig{fig:post-processing-pipelineX} have been reported, e.g.\  by the USCB for its TopDown algorithm \cite{Abowd2021, CommitteeOnNationalStatistics2020}. Their pipeline starts with the Census microdata from which \qm{noisy measurements} are made to create histograms by applying differential privacy. A post-processing step follows to satisfy invariants and constraints and to generate privacy-protected microdata, which is then used to create the USCB products \cite{CommitteeOnNationalStatistics2020}. In particular, this pipeline resembles that for the noisy \ArgMax\ from Sec. \ref{a1313}. According to the USCB, from the two primary sources of error within the TopDown algorithm, i.e. \qm{measurement error} from adding noise into histograms cells counts and \qm{post-processing error}, the latter is \qm{much larger} \cite{CommitteeOnNationalStatistics2020}.

Quantitative Information Flow was pioneered by Clark, Hunt, and Malacaria \cite{Clark:01:ENTCS}, followed by a growing community, e.g. \cite{Malacaria:07:POPL,Chatzikokolakis:08,Smith2009}, and its principles have been organized in~\cite{Alvim20:Book}. Moreover, the Haskell-based Kuifje programming language \cite{Gibbons2020,Bognar2019} is able to interpret exactly the code given in the algorithms provided here.

\section{Conclusions}
We have presented a novel formal analysis technique for investigating the fine-grained behaviour of utility in privacy-preserving pipelines for large datasets. The analysis is based on Kronecker products and the  theory of quantitative information flow,
including its refinement order and the soundess/completeness property.
Using this framework we are able to define the property of stability which seems to be a useful feature for simplifying the goal of finding the smallest privacy parameter to achieve a desired level of utility. Further, we are able to formally verify that any pipeline design using random response and counting will certainly be stable. Moreover we have also been able to explain anomalous behaviour observed in other privacy-preserving mechanisms that use geometric perturbers and counting. In the future a technique that produces small counterexamples might prove useful as a debugging tool.

One of our most compelling contributions is the use of Kronecker products applied to large data analyses, offering the prospect of tractable formal analysis for the verification of fine-grained statistical properties, perhaps applied to the verification of implementations of privacy-preserving libraries. In the future we would like to further develop the Kronecker method to complement the proof of privacy and utility in complex privacy pipelines.

\subsection*{Acknowledgments}

M\'{a}rio S.\ Alvim and Gabriel H.\  Nunes were partially supported by CNPq, CAPES, and FAPEMIG.
The research was partially funded by the European Research Council (ERC) project Hypatia, grant agreement № 835294.

\bibliographystyle{ieeetr}
\balance
\bibliography{postprocessing}

\onecolumn

\section*{Appendices}
\appendix
\label{appendices}

\section{Proofs for \Sec{s1347X}}\label{a1154}
\begin{lemma} (used to prove ($\J\dagger$) in \Lem{t1201X})
Let $C$ be a $2{\times}2$ stochastic matrix (i.e.\ a channel) and $Y$ a tally as defined above. Then we have
\begin{equation}
Y{\MM}\YI\MM \nfold{N}{C}\MM Y \Wide{=}  \nfold{N}{C}\MM Y\Q.
\end{equation}
\begin{proof}
Observe first that $Y_{s,k} = Y_{\sigma(s), k}$ for all permutations $\sigma$ of the indices of tuple $s$ in $\bool^N$. Further, since $Y$ is deterministic, there is exactly one value of $k$  for which that element of $Y$ is $1$.  From this, we can define an equivalence relation $(\sim)$ on $N$-tuples $s,s'$ in $\bool^N$, namely
\[
	s\sim s' \WIDERM{iff} Y_{s,-}=Y_{s',-}\Q.
\]
Observe that each equivalence class is generated by all permutations of any element from the class: in effect two tuples are equivalent just when they yield the same $Y$-count.

For example if $s= (\true, \false, \true)$ then it would have $Y$-count of $2$, so that $Y_{s,2}=1$ and $Y_{s, k}=0$ for $k\neq2$.

\bigskip
As a post-processor following $\nfold{N}{C}$, the tally $Y$ merges all columns according to their equivalence classes.

As a  pre-processor for $ \nfold{N}{C}\MM Y$, any fixed left-inverse $\YI$ of the tally removes all but one of the rows corresponding to each equivalence class of tuple --- and then the following tally $Y$ replaces the removed rows with the remaining row in the equivalence class.

\medskip
The above taken together means that \Eqn{e1016-a} holds if and only if $~Y_{s,-}= (\nfold{N}{C}\MM Y)_{\sigma(s),-}~$ for all permutations $\sigma$.
And we can show that
by direct calculation, since $C$ is so specific: let it be the matrix
\[
\left(\begin{array}{cc}
a & b\\
c& d
\end{array}\right)\Q,
\]
and let $Y$ be such that $Y_{s, k} = 1$, i.e.\ that tuple $s$ contains exactly $k$ occurrences of $\true$. Then, from a direct calculation we have that
$~(\nfold{N}{C}\MM Y)_{s,m}~$ is the sum of the  probabilities that the output tuple $s'$ from $\nfold{N}{C}$, i.e.\ the tuple passed from $C$ to $Y$, has $m$ occurrences of $\true$. The overall probability distribution of the output tuple is determined by executing $k$ (independent) random choices according to the distribution given by the first row of $C$, and $N{-}k$ (independent) random choices according to the distribution given by the second row of $C$. And to find that distribution
we need to sum for each $m$ the probabilities of all the output sequences that have tally $m$.

Using probability-generating functions, that is equal to
the $m$'th coefficient of $t^m$ in the polynomial $(at + b)^k(ct + d)^{N-k}$. Since this is not affected by permutations of $s$, the result follows.
\end{proof}
\end{lemma}

\section{Stability proof for a Geometric perturber and counting outliers}\Label{AS1553}
Recall from \Sec{s1215X} the scenario where the analyst receives perturbed data according to a geometric definition with three possible outputs. Here she wants to  count outliers \emph{of either kind} thus uses an aggregator as in \Def{s1123}:
\[
L_{y, t}  \Wide{\Defs} ~~ [y\in \{0, 2\}]~~~~\textit{\quad and\quad}~~~~\quad L_{y, f}  \Wide{\Defs} ~~ [y\not\in \{0, 2\}]~.
\]
  In this case, it turns out that $L$ and $\nfold{N}{L}$ satisfy the conditions for \Thm{t1321} wrt.\ the witness for refinement for $G_3 \ChanRef G_2$, thus we do have that $\nfold{N}{G_3}\MM (\nfold{N}{L}\MM Y) \ChanRef \nfold{N}{G_2}\MM  (\nfold{N}{L}\MM Y)$. The details are as follows:

\begin{enumerate}
\item The witness $W$ for refinement $G_3 \ChanRef G_2$ is given by $(G_3)^{-1}\MM G_2$ which is:
\[
\begin{bmatrix}
               5/6 & 1/12 & 1/12 \\
               7/30 & 8/15 & 7/30 \\
               1/12 & 1/12 & 5/6
            \end{bmatrix}
\]
\item The full matrix form of $L$ above is:
\[
\begin{bmatrix}
               1 & 0 \\
               0 & 1 \\
               1 & 0
            \end{bmatrix}~,
\]
\item and its left inverse $\LI$ is:
\[
\begin{bmatrix}
               1 & 0 &0\\
               0 & 1 & 0
            \end{bmatrix}~
\]
\item We can now readily confirm that $L \MM \LI \MM W \MM L = W \MM L$~, as required.

\end{enumerate}

\section{More details for \ArgMax}\Label{AS1623}

Kronecker products satisfy the pleasant additive property for privacy, which can be conveniently expressed using $d$-privacy \cite{DBLP:phd/hal/Fernandes21}. Let $d$ be some metric over the inputs of a channel $C$. This means that for any inputs $x,x'$, and ouput $y$, we must have that $C_{x,y}\leq e^{d(x,x')}{\times}C_{x', y}$ Then we can say:

\begin{quotation}
If $C$ is $d$-private, then $\nfold{N}{C}$ is $d_M$-private~,
\end{quotation}
where $d_M$ is the ``Manhattan'' distance on the $N$-sequence inputs to  $\nfold{N}{C}$, that is $d_M(v, v')= \sum_{0{\leq} i {<} N} d(v_i, v_i')$.

We can apply this to the noisy \ArgMax\ by observing that $\GD$ is $\VE{\times}{|\cdot|}$-private, where the metric $|\cdot|$ is simply the difference $|x-x'|$ for integer inputs $x, x'$.
Thus $\nfold{N}{\GD}$ is $\VE{\times}|\cdot|_M$-private.

Post-processing by \ArgMax\ certainly preserves this privacy property, by \DPI, thus noisy \ArgMax~ is also  $\VE{\times}|\cdot|_M$-private. But this reduces to the normal meaning of $\VE$-privacy because we only need to look at two \emph{adjacent} histograms, namely inputs $h, h'$ that differ in exactly one bar (or index), and only by one value, \emph{viz.}\ there is some $j$ such that $|h_j-h'_j|=1$, and for all other histogram bars $i\neq j $ we have that $h_i=h'_i$. In this case we have that $\VE{\times}|h-h'|_M= \sum_{0{\leq} i {<} N} \VE{\times}|h_i-h'_i|= \VE$. The result now follows.

\section{Examples of instability}\Label{appendix:instability-examples}
\input{appendix-instability-examples.tex}

\section{Counter-example for sum queries}\label{a1531}
\Af{Is this from Annabelle? No.}

For the sum query counter-example, we had:

\[
     R_{2}
     = \begin{bmatrix}
               1/2 & 1/4 & 1/4 \\
               1/4 & 1/2 & 1/4 \\
               1/4 & 1/4 & 1/2
            \end{bmatrix}
       \qquad
	 R_{3}
     = \begin{bmatrix}
             3/5 & 1/5 & 1/5 \\
             1/5 & 3/5 & 1/5 \\
             1/5 & 1/5 & 3/5
           \end{bmatrix}
\]

And the sum query is defined:

\[
   \centering
    S 
    = \begin{tabular}{c | c c c c c}
     & 0 & 1 & 2 & 3 & 4 \\
    \hline
    00 & 1 & 0 & 0 & 0 & 0 \\
    01 & 0 & 1 & 0 & 0 & 0 \\
    02 & 0 & 0 & 1 & 0 & 0 \\
    10 & 0 & 1 & 0 & 0 & 0 \\
    11 & 0 & 0 & 1 & 0 & 0 \\
    12 & 0 & 0 & 0 & 1 & 0 \\
    20 & 0 & 0 & 1 & 0 & 0 \\
    21 & 0 & 0 & 0 & 1 & 0 \\
    22 & 0 & 0 & 0 & 0 & 1
    \end{tabular}
\]

Now, we want to show that 
$(R_{3} \kron R_{3}) \cdot S \not\ChanRef (R_{2} \kron R_{2}) \cdot S$.
We can compute that 
$(R_{3} \kron R_{3}) \cdot S$ 
is:

\[
  \centering
  \begin{tabular}{c | c c c c c }
  & 0 & 1 & 2 & 3 & 4 \\
  \hline
  00 & 0.36 & 0.24 & 0.28 & 0.08 & 0.04 \\
01 & 0.12 & 0.4  & 0.28 & 0.16 & 0.04 \\
02 & 0.12 & 0.16 & 0.44 & 0.16 & 0.12 \\
10 & 0.12 & 0.4  & 0.28 & 0.16 & 0.04 \\
11 & 0.04 & 0.24 & 0.44 & 0.24 & 0.04 \\
12 & 0.04 & 0.16 & 0.28 & 0.4   & 0.12 \\
20 & 0.12 & 0.16 & 0.44 & 0.16 & 0.12 \\
21 & 0.04 & 0.16 & 0.28 & 0.4  & 0.12 \\
22 & 0.04 & 0.08 & 0.28 & 0.24 & 0.36 \\
  \end{tabular}
\]

and 
$(R_{2} \kron R_{2}) \cdot S$
is:

\[
  \centering
  \begin{tabular}{c | c c c c c }
  & 0 & 1 & 2 & 3 & 4 \\
  \hline
00 & 0.25 & 0.25 & 0.3125 & 0.125 & 0.0625 \\
01 & 0.125 & 0.3125 & 0.3125 & 0.1875 & 0.0625 \\
02 & 0.125 & 0.1875 & 0.375 & 0.1875 & 0.125 \\
10 & 0.125 & 0.3125 & 0.3125 & 0.1875 & 0.0625 \\
11 & 0.0625 & 0.25 & 0.375 & 0.25 & 0.0625 \\
12 & 0.0625 & 0.1875 & 0.3125 & 0.3125 & 0.125 \\
20 & 0.125 & 0.1875 & 0.375 & 0.1875 & 0.125 \\
21 & 0.0625 & 0.1875 & 0.3125 & 0.3125 & 0.125 \\
22 & 0.0625 & 0.125 & 0.3125 & 0.25 & 0.25 \\
  \end{tabular}
\]

And we find the following counter-example loss function, in which columns correspond to actions and rows are $x$-labelled:
\Cf{\NStuff{Natasha}, maybe play with this loss function by using say only 2 decimal places. I think in our current framework the rows of the loss function are \XX-labelled and the columns are \WW-labelled.}

\[
  \centering
  \begin{tabular}{c | c c c c c }
  & $w_1$ & $w_2$ & $w_3$ & $w_4$ & $w_5$ \\
  \hline
 00 &  0.33525164 & 0.67014371 & 0.50295689 & 0.64597016 & 0.34089403 \\
01 & 0.58508164 & 0.41408977 & 0.50010053 & 0.42214762 & 0.57379685 \\
02 & 0.74867853 & 0.25036935 & 0.49712065 & 0.25036935 & 0.74867853 \\
10 & 0.58508164 & 0.41408977 & 0.50010053 & 0.42214762 & 0.57379685 \\
11 & 0. & 1. & 0.49521641 & 1. & 0. \\
12 & 0.57379685 & 0.42214762 & 0.50010053 & 0.41408977 & 0.58508164 \\
20 & 0.74867853 & 0.25036935 & 0.49712065 & 0.25036935 & 0.74867853 \\
21 & 0.57379685 & 0.42214762 & 0.50010053 & 0.41408977 & 0.58508164 \\
22 & 0.34089403 & 0.64597016 & 0.50295689 & 0.67014371 & 0.33525164 \\
  \end{tabular}
\]

\section{Example: direct proof of stability for a special case}\Label{s1028W}

We illustrate the stability proof in \Sec{s1127} (random response and counting) by considering the special case in which there are  $N{=}3$ participants and a single choice $k$ in \KK\ is tallied. (In \Sec{s1127} any $1{\leq}N$ of participants was allowed, and any subset of choices in \KK\ could be tallied.)

\smallskip
From \Lem{l1601X} we recall that any single-respondent $\RR^K$ can be written $a\Id+b\Uf$, where \Id\ is the identity matrix and \Uf\ is a matrix all of whose elements are \NF{1}{K}, both matrices being of size $K{\times}K$. In general however we're interested in the $N{>}1$ -respondent case, with $\nfold{N}{(\RR^K)}$ as the perturber: it would then be
\begin{equation}\label{e1913W}
	\nfold{N}{(\RR^K)} \Wide{=} \underbrace{(a\Id+b\Uf) \kron \cdots \kron (a\Id+b\Uf)}_N \Q,
	\vspace{-2ex}
\end{equation}
for some $a,b$.

\bigskip\noindent Here we specialise to $N{=}3$, that is $3$ respondents, and we use the distribution laws for $\kron$ to see that \Eqn{e1913W}, that is \quad$\nfold{3}{(\RR^K)}_{(x_1,x_2,x_3),(y_1,y_2,y_3)}$\quad becomes
\newcommand\zS[1]{\parbox{4em}{$#1\,\times$}}
\begin{Reason}
\Step{}   {\zS{a^3}  (\Id\kron\Id\kron\Id)_{(x_1,x_2,x_3),(y_1,y_2,y_3)}}
\Step{$+$}{\zS{a^2b} (\Id\kron\Id\kron\Uf)_{(x_1,x_2,x_3),(y_1,y_2,y_3)}}
\Step{$+$}{\zS{a^2b} (\Id\kron\Uf\kron\Id)_{(x_1,x_2,x_3),(y_1,y_2,y_3)}}
\Step{$+$}{\zS{a^2b} (\Id\kron\Uf\kron\Uf)_{(x_1,x_2,x_3),(y_1,y_2,y_3)}}
\Step{$+$}{\zS{ab^2} (\Uf\kron\Id\kron\Id)_{(x_1,x_2,x_3),(y_1,y_2,y_3)}}
\Step{$+$}{\zS{ab^2} (\Uf\kron\Id\kron\Uf)_{(x_1,x_2,x_3),(y_1,y_2,y_3)}}
\Step{$+$}{\zS{ab^2} (\Uf\kron\Uf\kron\Id)_{(x_1,x_2,x_3),(y_1,y_2,y_3)}}
\Step{$+$}{\zS{b^3}  (\Uf\kron\Uf\kron\Uf)_{(x_1,x_2,x_3),(y_1,y_2,y_3)}\Q,}
\end{Reason}
where each $x_n$ (for $n{=}1,2,3$) is the choice in \KK\ made by respondent $n$, and each $y_n$ is the output to which the (individual) {\RR}\ has perturbed the input choice $x_n$.

Evaluating each summand of the above (using \Def{d1130X} of Kroneckering) gives us
\begin{equation}\label{e1525Y}
\begin{minipage}{0.4\textwidth}
\renewcommand\zS[1]{\parbox{5em}{$#1\,\times$}}
\begin{Reason}
\Step{}   {\zS{a^3}      \Iv{x_1{=}y_1\land x_2{=}y_2\land x_3{=}y_3}}
\Step{$+$}{\zS{a^2b/K}   \Iv{x_1{=}y_1\land x_2{=}y_2}}
\Step{$+$}{\zS{a^2b/K}   \Iv{x_1{=}y_1\land x_3{=}y_3}}
\Step{$+$}{\zS{a^2b/K}   \Iv{x_2{=}y_2\land x_3{=}y_3}}
\Step{$+$}{\zS{ab^2/K^2} \Iv{x_1{=}y_1}}
\Step{$+$}{\zS{ab^2/K^2} \Iv{x_2{=}y_2}}
\Step{$+$}{\zS{ab^2/K^2} \Iv{x_3{=}y_3}}
\Step{$+$}{\zS{b^3/K^3}  1\Q,}
\end{Reason}
where we use ``Iverson brackets'' $\Iv{-}$ \\ to convert Booleans to 0,1.
\end{minipage}
\end{equation}
\EQN{e1525Y} is a single real number: it is the element of \nfold{3}{(\RR^K)}\!\! indexed by row $(x_1,x_2,x_3)$ and column $(y_1,y_2,y_3)$.
Now we want to look at the details of how a counting query $T$ interacts with a $W$ of the general form \Eqn{e1525Y} above.
We want to know in particular when the equality \Eqn{e1914} from \Thm{t1321} holds, which we repeat here for our counting query $T$ specifically:
\begin{equation}\label{e1156XY}
	T{\MM}\TI\MM W\MM T \Wide{=} W\MM T \Q.
\end{equation}
Without loss of generality we assume that \KK\ is $\{1..K\}$, and we specialise to the case that $T$ is counting the single response $k{=}1$. (The Boolean aggregator $B$ from \Def{d1517-z} is more general because it can send \emph{any} fixed subset of \KK\ to \true.)

We begin with the leading $T{\MM}\TI$, i.e.\ on the left: its effect on an input 3-tuple $x$ is
simply to move it to some (possibly other) 3-tuple $x'$ in which the same number of 1's occur.

Next we look at the effect of the trailing $T$, which is on both sides of the equation: it simply groups together $y$-output 3-tuples that have the same number of 1's.

So to show that \Eqn{e1156XY} holds, we must argue that if we change an input $x$ to another $x'$ with the same number of 1's, then for \emph{any} number $0{\leq}t{\leq}N$ the sum \Eqn{e1525Y} does not change when taken over all output-tuples with $t$ occurrences of 1. (In this case $N{=}3$, but the argument will be given in terms of $N$.)

We make that argument by changing $x$ into $x'$ taking ``one step at a time''. Remember that ``1'' is the value in $1..K$ that we are counting
and assume --for the moment-- that all 1's in the input-tuple $x$ occur together at its beginning. That is, the input-tuple $x$ and the ``one step away'' $x'$ are both of the form $(1,\ldots,1,k',k''\ldots k''')$,
where all the primed $k$'s are not 1 (but they are not necessarily all different).

Now suppose that in a single step we change some element $x_n$ of $x$ to $x'_n$, i.e.\ we change the value at index $1{\leq}n{\leq}N$ that was supplied by Respondent $n$. We note first that neither $x_n$ nor $x'_n$ can be 1, because otherwise that step would change the number of 1's in $x$.

For each $y$ with a fixed count $0{\leq}t{\leq}N$ of 1's there are now three possibilities for each of its elements $y_n$:
\begin{enumerate}
\item If $y_n{=}1$, then none of the summands in \Eqn{e1525Y} will be affected, since neither $x_n$ nor $x'_n$ is 1.
\item If $y_n\,{=}\,x_n$ but $y_n\,{\neq}\,x'_n$, then for that $y$ some summands in \Eqn{e1525Y} that include conjunct $x_n{=}y_n$ will change from 1 to 0, because the conjunct is no longer true.\,%
\footnote{The summands that \emph{don't change} will be 0 already because of some other conjunct.}
But there is always \emph{another} $y'$ with the same count $t$ of 1's, exactly identical to $y$ except that $y'_n{=}x'_n$, in which exactly the opposite change (i.e.\ from 0 to 1) will occur.
Thus the overall sum --over \emph{all} $y$'s with that count $t$ of 1's-- remains the same.
\item The third case, where $y_n{\neq}x_n$ but now $y_n{=}x'_n$, is analogous to the second.
\end{enumerate}

Finally, we can relax the restriction that all of the $x$-tuple's occurrences of 1 must come first: that is because any target $x'$ where the some 1's do \emph{not} come first can be reached from the $x$ with all its 1's first simply by following the above step-by-step procedure until an intermediate $x''$ is reached that (still) has all its 1's first \emph{and} at the same time is a permutation of the final $x'$ desired. It will have the same \Eqn{e1525Y}-sum for any $y$ having $t$ instances of 1 as $x$ did, from the step-by-step argument above; but it will also have the same sum as the actual target $x'$, because the subset of \YY\ having exactly $t$ instances of 1 is permutation closed.

That completes our ``proof by example'' of \Eqn{e1156XY} and hence (re-using material from \Sec{s1127}) that multiple-participant random-response folowed by a counting query is stable.

\end{document}

%% file: macros.tex
\newtheorem{theorem}{Theorem}
\newenvironment{Proof}{\\[.5ex]{\bf Proof:}~~\bgroup\rm}{\egroup}
\newtheorem{definition}[theorem]{Definition}
\newtheorem{proposition}[theorem]{Proposition}
\newtheorem{lemma}[theorem]{Lemma}
\newtheorem{corollary}[theorem]{Corollary}

\newtheorem{remark}[theorem]{Remark}

\newcommand\In {{:}\;}
\newcommand\Wide[1] {\quad#1\quad}
\newcommand\WideRM[1] {\quad{\rm #1}\quad}
\newcommand\WIDERM[1] {\qquad{\rm #1}\qquad}
\newcommand\Defs {:=}
\newcommand\DefsR {:=\quad}

\newcommand\UU {\ensuremath{U}}
\newcommand\Util[3]	{\UU_{#1}(#2, #3)}

\newcommand\Dist {\mathbb{D}}

\newcommand\QIF {\textit{QIF}}
\newcommand\DPI {\textit{DPI}}

\newcommand\ChanRef {\ensuremath{\sqsubseteq}}
\newcommand\CR {\,{\ChanRef}\,}

\newcommand\ChanRefN {\ensuremath{\not\sqsubseteq}}
\newcommand\ChanUt {\ensuremath{\preceq}}

\newcommand\Eqn[1] {(\ref{#1})}
\newcommand\EQn[1] {Eqn.\,(\ref{#1})}
\newcommand\EQN[1] {Equation\,(\ref{#1})}
\newcommand\Itm[1] {(\ref{#1})}
\newcommand\Thm[1] {Thm.\,\ref{#1}}
\newcommand\Lem[1] {Lem.\,\ref{#1}}

\newcommand\Fig[1] {Fig.\,\ref{#1}}
\newcommand\Def[1] {Def.\,\ref{#1}}

\newcommand\Alg[1] {Alg.\,\ref{#1}}

\newcommand\Cor[1] {Cor.\,\ref{#1}}

\newcommand\calx {\mathcal{X}}
\newcommand\caly {\mathcal{Y}}

\newcommand\calw {\mathcal{W}}

\newcommand\kron {\otimes}
\newcommand\nfold[2] {\ensuremath{{#2}^{\kron #1}}}

\newcommand\Sec[1] {Sec.\,\ref{#1}}

\newcommand\App[1] {App.\,\ref{#1}}

\newenvironment{Reason}{\vspace{-.0em}\begin{tabbing}\hspace{3em}\= \hspace{1cm} \= \kill}
    {\end{tabbing}\vspace{-1em}}
\newcommand\Space {~\\}
\newcommand\Step[2] {#1 \> $\begin{array}[t]{@{}llll}#2\end{array}$ \\}
\newcommand\StepR[3] {#1 \> $\begin{array}[t]{@{}llll}#3\end{array}$
    \` {\RF \makebox[0pt][r]{\small\begin{tabular}[t]{r}``#2''\end{tabular}}} \\}
\newcommand\WideStepR[3] {#1 \>
    $\begin{array}[t]{@{}ll}~\\#3\end{array}$ \`
    {\RF \makebox[0pt][r]{\begin{tabular}[t]{r}``#2''\end{tabular}}} \\}
\newcommand\RF {\small}

\newcommand{\qm}[1]{``#1''}

\newcommand\FFF {\ensuremath{\mathbb F}}
\newcommand\RRRR {\ensuremath{\mathbb R \mathbb R}}

\newcommand\JJ {\hspace{.2em}} 
\newcommand\BBI {\ensuremath{^{-1}\hspace{-.15em} (\nfold{N}{B}\MM Y) }}
\newcommand\YI {\ensuremath{^{-1}\hspace{-0.05em}Y}}

\newcommand\Id {\ensuremath{\mathbb I}}
\newcommand\Uf {\ensuremath{\mathbb U}}
\newcommand\Iv[1] {[#1]}
\newcommand\Cite[1] {{\textcolor{red}{[#1]}}}
\newcommand\XX {\ensuremath{\mathcal X}}
\newcommand\KK {\ensuremath{\mathcal K}}
\newcommand\YY {\ensuremath{\mathcal Y}}
\newcommand\ZZ {\ensuremath{\mathcal Z}}

\newcommand\WW {\ensuremath{\mathcal W}}
\newcommand\WWW {\ensuremath{\mathbb W}}

\newcommand\DP {\textit{DP}}
\newcommand\EDP {\VE-\textit{DP}}

\newcommand\VE {\ensuremath{\varepsilon}}
\newcommand\RR[1][] {\ensuremath{\ifx\empty#1\empty{\mathit RR}_\VE\else {\mathit RR}_{#1}\fi}}
\newcommand\RRX {\RR[\relax]}
\newcommand\GD[1][] {\ensuremath{\ifx\empty#1\empty{\mathit GD}_\VE\else {\mathit GD}_{#1}\fi}}
\newcommand\GDX {\GD[\relax]}

\newcommand\MM {\ensuremath{\,{\cdot}\,}} 
\newcommand\Fun {{\rightarrow}}
\newcommand\MFun {\mathbin{\rightarrowtriangle}}
\newcommand\Nat {\ensuremath{\naturals}}
\newcommand\Real {\ensuremath{\reals}}

\newcommand\NF[2] {\ensuremath{\nicefrac{#1}{#2}}}
\newcommand\WIDE[1] {\quad\quad#1\quad\quad}
\newcommand\LF {\ensuremath{\ell}}
\newcommand\ArgMax {\ensuremath{\mathit ArgMax}}
\newcommand\ArgMaxI {\ensuremath{^{-1}\hspace{-.15em}\mathit ArgMax}}
\newcommand\LFU {\ensuremath{{\hat\ell}}} 
\newenvironment{Theorem}{\begin{theorem}}{\hfill$\square$\end{theorem}}
\newenvironment{Definition}{\begin{definition}}{\hfill$\square$\end{definition}}

\newenvironment{Lemma}{\begin{lemma}}{\hfill$\square$\end{lemma}}
\newcommand\COMPAS {{\tt COMPAS}}
\newcommand\Label[1] {\label{#1}\marginpar{\tiny\color{gray}\tt#1}}
\newcommand\Par {\newline\hspace*{1em}}

\newcommand\Yes {{\tt yes}}
\newcommand\No {{\tt no}}

\newcommand\Q[1] {\makebox[0pt][l]{\quad\mbox{#1}}}
\newcommand\Matrix[2] {\ensuremath{\Real(#1{\MFun}#2)}}
\newcommand\CCC {\ensuremath{\mathbb C}}
\newcommand\RRR {\ensuremath{\mathbb R}}
\newcommand\TTT {\ensuremath{\mathbb T}}
\newcommand\PPP {\ensuremath{\mathbb P}}

\newcommand\PI {\ensuremath{^{-1}\hspace{-.15em}P}}

\newcommand\TI {\ensuremath{^{-1}\hspace{-.15em}T}}
\newcommand\LI {\ensuremath{^{-1}\hspace{-.15em}L}}
\newcommand\Pn {\ensuremath{\overline{p}}}
\newcommand\Kk {\ensuremath{\hat{k}}}
\newcommand\Kn {\ensuremath{\overline{k}}}

\newcommand\YCells {\ensuremath{\widehat Y}}
\newcommand\YCell {\ensuremath{\widehat y}}
\newcommand\XCell {\ensuremath{\widehat x}}

\newcommand\Grey {\color{gray}}
\newcommand\J {\hspace{.1em}} 
\newcommand\RHS {\textit{rhs}}
\newcommand\LHS {\textit{lhs}}

\newcommand\BB {\ensuremath{^{-1}\hspace{-.15em}B}}

\newcommand\KKR {\ensuremath{^{-1}\hspace{-.15em} (\nfold{N}{B}) }}

\newcommand{\domainofdatasets}{\cald}

\newcommand{\attrset}{\cala}

\newcommand{\domainofrows}[1]{\textit{rows}(#1)}
\newcommand{\rowprior}{\pi^{\star}}

\newcommand{\senscol}{a_{s}}
\newcommand{\sensset}{\domain{\senscol}}

\newcommand{\usefulcol}{a_{u}}

\newcommand{\countquery}[1]{\texttt{count}_{q}}
\newcommand{\countprop}{\countquery{\prop}}

\newcommand{\utility}{\textit{utility}}

\newcommand{\range}[1]{\textit{range}(#1)}

\newcommand{\context}{\Gamma}
\newcommand{\contextobv}{\Gamma^{\textit{obv}}}
\newcommand{\contextloc}{\Gamma^{\textit{loc}}}

\newcommand{\cala}{\mathcal{A}}

\newcommand{\cald}{\mathcal{D}}

\newcommand{\cals}{\mathcal{S}}

\newcommand{\dist}{\mathbb{D}}

\newcommand*{\ldblbrace}{\{\mskip-5mu\{}
\newcommand*{\rdblbrace}{\}\mskip-5mu\}}
\newcommand{\multiset}[1]{\ldblbrace#1\rdblbrace}

\newcommand{\domain}[1]{\mathit{domain}(#1)}

\newcommand{\bool}{\texttt{Bool}}
\newcommand{\true}{\texttt{true}}
\newcommand{\false}{\texttt{false}}

\newcommand{\naturals}{\mathbb{N}}
\newcommand{\reals}{\mathbb{R}}

\newcommand{\eqdef}{\stackrel{\text{def}}{=}}

\newcommand{\mech}{M_{\countprop}}

\newcommand{\mechloc}{\mech^{\textit{loc}}}
\newcommand{\rand}{R}

\newcommand{\randloc}{\rand^{\textit{loc}}}
\newcommand{\prop}{q}

\newcommand{\distcontext}{p^{\context}}
\newcommand{\distobv}{p^{\contextobv}}
\newcommand{\distloc}{p^{\contextloc}}

\newcommand{\dirac}[2]{\delta_{#1}(#2)}

\newcommand{\xstar}{x^{\star}}
\newcommand{\extgen}[2]{#1{\cup}#2}
\newcommand{\extd}[1]{\extgen{D}{#1}}
\newcommand{\dstar}{\extd{\xstar}}
\newcommand{\dx}{\extd{x}}
\newcommand{\dy}{\extgen{d}{y}}

\newcommand{\indu}{w}



%% file: appendix-instability-examples.tex
As we discussed in the main body of this paper, ``instability''
is a situation in which privacy is increased
(resp.\ decreased) in part of a data-release pipeline
(and, as a consequence of differential privacy's property of resistance to post-processing, in the overall pipeline
as well), and yet the utility of the whole pipeline
also increases (resp.\ decreases).
The phenomenon is perhaps surprising because privacy
practitioners normally make decisions under the assumption that privacy and utility
go in opposite directions.

In this section we provide two examples of instability.
The first is derived from a common data-release pipeline
applied to a real dataset, and it illustrates how the
phenomenon can indeed occur in practical scenarios (\Sec{ss1554}).
The second is a minimal example of that same phenomenon,
but now one that can be checked step-by-step, therefore
certifying that the instability we discovered did not arise from
coding-- or rounding-errors.

\subsection{Instability that can be checked manually}\Label{s1102X}

The instability from Sec.~\ref{s0959X} above
was calculated by the authors running their own code.
(It was the database itself that was public.)
\Mf{Clarify here that we run  ``their'' (our) code, since the instability was not (I think) reported in the QONFEST paper. \GStuff{Gabriel?} \NStuff{Natasha?}}
We emphasize that the results are not limited to a few runs of the
mechanism, but it is an exact computation of the expected behavior of
the mechanism (in terms of \VE's and \UU's) with respect to all coin
tosses of the mechanism itself and the uncertainty of the
data analyst's prior distribution on the input to the mechanism.

In order to verify that the phenomenon is not a consequence
of incorrect code or rounding errors,
we constructed a small scenario
of instability that can be checked by hand and employs only exact, rational arithmetic.
In this example we have that:
\begin{enumerate}
	\item There are only 4 individuals, each with a sensitive
	attribute taking an integer value between 0 and 6.%
	\item The data analyst  knows the real values of the
	first 3 individuals	to be $(0,1,1)$, and has a uniform prior on the
	value of the 4\textsuperscript{th} individual, that is,
	$\NF{1}{7}$ for each of the values from 0 to 6.
	\item The data analyst runs the the query
	\emph{``\J How many individuals have sensitive value 0 in the dataset?\J''}, and her
	loss function is based on the absolute difference between the real value $x$
	for the 4\textsuperscript{th} individual and the value $w$ she guesses for it,
	that is we define \LF\ as
	\begin{equation}\label{e1842}
		\LF(x,w) \Wide{=} 1000 * |w - x|~\Q,
	\end{equation}
	which can be represented in matrix form (with guesses $w$ as rows and real values $x$ as columns) as
	$$
	\begin{array}{c|ccccccc}
		& 0  &  1 &  2 &  3 &  4 &  5 &  6 \\ \hline
		0 & 0  & 1000 & 2000 & 3000 & 4000 & 5000 & 6000 \\
		1 & 1000 & 0  & 1000 & 2000 & 3000 & 4000 & 5000 \\
		2 & 2000 & 1000 & 0  & 1000 & 2000 & 3000 & 4000 \\
		3 & 3000 & 2000 & 1000 & 0  & 1000 & 2000 & 3000 \\
		4 & 4000 & 3000 & 2000 & 1000 & 0  & 1000 & 2000 \\
		5 & 5000 & 4000 & 3000 & 2000 & 1000 & 0  & 1000 \\
		6 & 6000 & 5000 & 4000 & 3000 & 2000 & 1000 & 0  \\
	\end{array}~.
	$$
	\item The truncated geometric mechanism \GD\ realizing \EDP\,%
	\footnote{That is, we use \GD\ for the perturber defined using the geometric probability distribution, tuned to realise \VE\ differential privacy.
		\CStuff{Mathematically speaking, this \GDX\ (i.e.\ without the \VE) is the function $f$ in \Def{d1322}.}
	}
	is then used for perturbation
	of each individual record, where in
	this first case $\VE$ is $~\ln 3.50$.
	If we represent \GD's effect as a matrix whose rows
	contain possible actual
	values for each individual, and
	whose columns contain
	possible randomized values,
	we get
\begin{align}
\label{e1843X2}
\begin{array}{c|ccccccc}
	& 0  &  1 &  2 &  3 &  4 &  5 &  6 \\ \hline
	0 & \nicefrac{7}{9} & \nicefrac{10}{63} & \nicefrac{20}{441} & \nicefrac{40}{3087} & \nicefrac{80}{21609} & \nicefrac{160}{151263} & \nicefrac{64}{151263} \\
	1 & \nicefrac{2}{9} & \nicefrac{5}{9} & \nicefrac{10}{63} & \nicefrac{20}{441} & \nicefrac{40}{3087} & \nicefrac{80}{21609} & \nicefrac{32}{21609} \\
	2 & \nicefrac{4}{63} & \nicefrac{10}{63} & \nicefrac{5}{9} & \nicefrac{10}{63} & \nicefrac{20}{441} & \nicefrac{40}{3087} & \nicefrac{16}{3087} \\
	3 & \nicefrac{8}{441} & \nicefrac{20}{441} & \nicefrac{10}{63} & \nicefrac{5}{9} & \nicefrac{10}{63} & \nicefrac{20}{441} & \nicefrac{8}{441} \\
	4 & \nicefrac{16}{3087} & \nicefrac{40}{3087} & \nicefrac{20}{441} & \nicefrac{10}{63} & \nicefrac{5}{9} & \nicefrac{10}{63} & \nicefrac{4}{63} \\
	5 & \nicefrac{32}{21609} & \nicefrac{80}{21609} & \nicefrac{40}{3087} & \nicefrac{20}{441} & \nicefrac{10}{63} & \nicefrac{5}{9} & \nicefrac{2}{9} \\
	6 & \nicefrac{64}{151263} & \nicefrac{160}{151263} & \nicefrac{80}{21609} & \nicefrac{40}{3087} & \nicefrac{20}{441} & \nicefrac{10}{63} & \nicefrac{7}{9} \\
\end{array}~.
\end{align}
	Now, if each individual randomises their own value with the matrix above, and then all perturbed values are shuffled to break the link between each data subject and their perturbed value, we
	obtain the following matrix that maps the value of the 4\textsuperscript{th}
	individual
	(rows)
	to the final count of the query
	(columns), given that the actual values for the first three individuals are known to be (0,1,1):
\begin{align}
\label{e1833X2}
\begin{array}{c|ccccc}
	& 0 & 1 & 2 & 3 & 4 \\ \hline
	0 & \nicefrac{196}{6561} & \nicefrac{1484}{6561} & \nicefrac{1067}{2187} & \nicefrac{1484}{6561} & \nicefrac{196}{6561} \\
	1 & \nicefrac{686}{6561} & \nicefrac{2989}{6561} & \nicefrac{742}{2187} & \nicefrac{604}{6561} & \nicefrac{56}{6561} \\
	2 & \nicefrac{826}{6561} & \nicefrac{3419}{6561} & \nicefrac{4544}{15309} & \nicefrac{2468}{45927} & \nicefrac{16}{6561} \\
	3 & \nicefrac{866}{6561} & \nicefrac{24793}{45927} & \nicefrac{30508}{107163} & \nicefrac{13756}{321489} & \nicefrac{32}{45927} \\
	4 & \nicefrac{6142}{45927} & \nicefrac{175271}{321489} & \nicefrac{210956}{750141} & \nicefrac{89252}{2250423} & \nicefrac{64}{321489} \\
	5 & \nicefrac{43154}{321489} & \nicefrac{1230337}{2250423} & \nicefrac{1471492}{5250987} & \nicefrac{610684}{15752961} & \nicefrac{128}{2250423} \\
	6 & \nicefrac{302398}{2250423} & \nicefrac{8619239}{15752961} & \nicefrac{10290044}{36756909} & \nicefrac{4246628}{110270727} & \nicefrac{256}{15752961} \\
\end{array}~.
\end{align}
	Under a uniform prior, that leads to an expected utility loss of
	$\nicefrac{1276949536000}{771895089} \approx 1654.3045$.

	\medskip
	\item If now, in the second case, \VE\ is set to the slightly higher
	(i.e.\ less private) value of $\ln 3.51$,
	we obtain instead the following perturbation matrix
	representing each individual's
	randomization via \GD:
\begin{align}
\label{e1844X2}
\begin{array}{c}
\begin{array}{c|cccc}
	& 0 & 1 & 2 & 3   \\ \hline
	0 & \nicefrac{351}{451} & \nicefrac{25100}{158301} & \nicefrac{2510000}{55563651} & \nicefrac{251000000}{19502841501}   \\
	1 & \nicefrac{100}{451} & \nicefrac{251}{451} & \nicefrac{25100}{158301} & \nicefrac{2510000}{55563651}   \\
	2 & \nicefrac{10000}{158301} & \nicefrac{25100}{158301} & \nicefrac{251}{451} & \nicefrac{25100}{158301}   \\
	3 & \nicefrac{1000000}{55563651} & \nicefrac{2510000}{55563651} & \nicefrac{25100}{158301} & \nicefrac{251}{451}   \\
	4 & \nicefrac{100000000}{19502841501} & \nicefrac{251000000}{19502841501} & \nicefrac{2510000}{55563651} & \nicefrac{25100}{158301}   \\
	5 & \nicefrac{10000000000}{6845497366851} & \nicefrac{25100000000}{6845497366851} & \nicefrac{251000000}{19502841501} & \nicefrac{2510000}{55563651}   \\
	6 & \nicefrac{1000000000000}{2402769575764701} & \nicefrac{2510000000000}{2402769575764701} & \nicefrac{25100000000}{6845497366851} & \nicefrac{251000000}{19502841501}
\end{array}
\\
\begin{array}{c|ccc}
	& 4 & 5 & 6 \\ \hline
	0 & \nicefrac{25100000000}{6845497366851} & \nicefrac{2510000000000}{2402769575764701} & \nicefrac{1000000000000}{2402769575764701} \\
	1 & \nicefrac{251000000}{19502841501} & \nicefrac{25100000000}{6845497366851} & \nicefrac{10000000000}{6845497366851} \\
	2 & \nicefrac{2510000}{55563651} & \nicefrac{251000000}{19502841501} & \nicefrac{100000000}{19502841501} \\
	3 & \nicefrac{25100}{158301} & \nicefrac{2510000}{55563651} & \nicefrac{1000000}{55563651} \\
	4 & \nicefrac{251}{451} & \nicefrac{25100}{158301} & \nicefrac{10000}{158301} \\
	5 & \nicefrac{25100}{158301} & \nicefrac{251}{451} & \nicefrac{100}{451} \\
	6 & \nicefrac{2510000}{55563651} & \nicefrac{25100}{158301} & \nicefrac{351}{451} \\
\end{array}
\end{array}~.
\end{align}
	This leads instead to the following mapping from the value of the 4\textsuperscript{th}
	individual (rows) to the final count
	of the query (columns),
	again given that perturbed values were shuffled and that the actual values for the first three individuals are known to be (0,1,1):

\begin{align}
\label{e1834X2}
\begin{array}{c}
\begin{array}{c|ccccc}
	& 0 & 1 & 2    \\ \hline
	0 & \nicefrac{1232010000}{41371966801} & \nicefrac{9350710200}{41371966801} & \nicefrac{20206526401}{41371966801}    \\
	1 & \nicefrac{4324355100}{41371966801} & \nicefrac{18874516401}{41371966801} & \nicefrac{14026065300}{41371966801}    \\
	2 & \nicefrac{5205365100}{41371966801} & \nicefrac{21587851501}{41371966801} & \nicefrac{4305102810200}{14521560347151}    \\
	3 & \nicefrac{5456365100}{41371966801} & \nicefrac{7848669386851}{14521560347151} & \nicefrac{1449286475370200}{5097067681850001}    \\
	4 & \nicefrac{1940284150100}{14521560347151} & \nicefrac{2782016305784701}{5097067681850001} & \nicefrac{502519091753940200}{1789070756329350351}    \\
	5 & \nicefrac{683549736685100}{5097067681850001} & \nicefrac{979201058430430051}{1789070756329350351} & \nicefrac{175766155095533010200}{627963835471601973201}    \\
	6 & \nicefrac{240176957576470100}{1789070756329350351} & \nicefrac{343970905019080947901}{627963835471601973201} & \nicefrac{61632115827522086580200}{220415306250532292593551}
\end{array}
\\
\begin{array}{c|cc}
	& 3 & 4 \\ \hline
	0 & \nicefrac{9350710200}{41371966801} & \nicefrac{1232010000}{41371966801} \\
	1 & \nicefrac{3796030000}{41371966801} & \nicefrac{351000000}{41371966801} \\
	2 & \nicefrac{776938510000}{14521560347151} & \nicefrac{100000000}{41371966801} \\
	3 & \nicefrac{217158615010000}{5097067681850001} & \nicefrac{10000000000}{14521560347151} \\
	4 & \nicefrac{70667993668510000}{1789070756329350351} & \nicefrac{1000000000000}{5097067681850001} \\
	5 & \nicefrac{24248997757647010000}{627963835471601973201} & \nicefrac{100000000000000}{1789070756329350351} \\
	6 & \nicefrac{8455851410934100510000}{220415306250532292593551} & \nicefrac{10000000000000000}{627963835471601973201}
\end{array}
\end{array}~.
\end{align}
	This counting query  \Eqn{e1834X2} above, under a uniform prior,
	has therefore \emph{not} led to a utility gain over \Eqn{e1833X2}, wrt.\ loss
	function \Eqn{e1842}.
	This happened even though the \EDP\
	of the \GDX-perturber increased between \Eqn{e1843X2} and \Eqn{e1844X2}, making the overall pipeline less private.
	Instead it led to a \emph{greater} loss
	(that is, smaller utility)
	of
	$\nicefrac{2552438666271995640492212000}{1542907143753726048154857} \approx 1654.3047$
	(i.e.\ \emph{more} than the loss of $1654.3045$ just above).
	This is exactly the phenomenon of instability.
\end{enumerate}

%
%

\subsection{Demonstrating instability \emph{without} a counter example}\Label{a1433X}
\Cf{I am still not sure I have got this section right.}
The \Sec{a1454X} illustrated by experiment the apparent stability of noisy \ArgMax\ in three scenarios. Is it still possible that there might be a prior distribution $\pi$ and a parameter \VE\ for which \GD\ followed by \ArgMax\ is unstable, and can we we show that to be so \emph{without} finding the actual $\pi$ and \VE?
\Cf{This is an informal section, trying to argue that we might be able to discover that {\GDX} plus \ArgMax\ is sometimes unstable (if indeed it is) using a \emph{standard} matrix package, rather than (e.g.)\ Kostas' tools. For the ``only if'' of \Thm{t1321} says that we can conclude (in general) $P \ChanRefN W\MM P$ from $P\MM\PI\MM W\MM P \neq W\MM P$\, (provided of course that \PI\ exists); and \emph{that} (since it involves only matrix multiplications) \emph{does not need Kostas}.}
What we want (for instability) is two geometric perturbers such that
\begin{equation}\label{e1446}
	\GD \ChanRef \GD[\VE']
	\WIDERM{and~yet}
	\GD\MM\ArgMax ~\ChanRefN~ \GD[\VE']\MM \ArgMax
	\Q,
\end{equation}
but in its Kroneckered version of course. The \LHS\ is equivalent to $\VE'{\leq}\VE$ because \GDX\ is a family. Can we get the $(\ChanRefN)$ on the \RHS\ from \Thm{t1321}?

A first step we can take is to note that $\GD \ChanRef \GD[\VE']$ means that $\GD\MM W = \GD[\VE']$ for some witness $W$,
and so the \RHS\ of \Eqn{e1446} requires
\begin{equation}\label{e1447}
	\GD\MM\ArgMax \Wide{\ChanRefN} \GD\MM W \MM \ArgMax
\end{equation}
for that $W$. Although it is tempting now to use the \textit{only if} of \Thm{t1321} with a $W$ satisfying
\[
	\ArgMax\MM\ArgMaxI\MM W\MM \ArgMax\Wide{\neq} W \MM \ArgMax
\]
to get  $\ArgMax\ChanRefN W \MM \ArgMax$, then inferring our target \Eqn{e1447} by applying $(\GD\MM)$ to both sides\ldots\ that's not valid. You cannot conclude $f(x)\,{\ChanRefN}\,f(x')$ from $x\,{\ChanRefN}\,x'$.

Instead, we ``start further back'' --- and it seems to depend on \GD's having an inverse.
\Cf{\Az Does this mean that this argument would work for \emph{any} perturber that has inverses? \AStuff{Yes I think so.} Should we say so?}
We assume $\GD\ChanRef\GD[\VE']$, the \LHS\ of \Eqn{e1446}, and then reason as follows (without Kronecker, because it can be added later
\Cf{I have temporarily forgotten where and how Kronecker is added. I \emph{think} the point is that with a single respondent, \ArgMax\ is trivial and so no instance of the assumption $(\dagger)$ will be found. I \emph{think} what I meant is that the reasoning below still works if you replace \GD\ by \nfold{N}{\GD}, and that with an $N{>}1$ one is more likely to find an instance of $(\dagger)$.}%
),
writing \ArgMax\ as $P$ for brevity:
\begin{Reason}
\WideStepR{}{assumption $(\dagger)$}{P\MM \PI\MM\GD^{-1}\MM\GD[\VE']\MM P
        \Wide{\neq}\GD^{-1}\MM\GD[\VE']\MM P}
\WideStepR{hence}{\EQN{e1505}, ``only if'', in \Thm{t1321}, with $W$ as $\GD^{-1}\MM\GD[\VE']$}
	{P\Wide{\ChanRefN}\GD^{-1}\MM\GD[\VE']\MM P}
\WideStepR{iff}{for all $V$, definition witness}
	{P\MM V \Wide{\neq} \GD^{-1}\MM\GD[\VE']\MM P}
\WideStepR{iff}{identity on left}
	{\GD^{-1}\MM\GD\MM P\MM V
	\Wide{\neq}
	\GD^{-1}\MM\GD[\VE']\MM P}
\WideStepR{hence}{in general $f(x){\neq}f(x')$ implies $x{\neq}x'$}
	{\GD\MM P\MM V
	\Wide{\neq}
	\GD[\VE']\MM P}
\WideStepR{hence}{using `for all $V$' from above, \Def{d1526X} of refinement and witness}
	{\GD\MM P
	\Wide{\ChanRefN}
	\GD[\VE']\MM P\Q,}
\end{Reason}
which is the \RHS\ of \Eqn{e1446} that we wanted. So \emph{if} we can find an instance of the assumption $(\dagger)$, now written out in full as
\begin{eqnarray}
	&&\ArgMax\MM\ArgMaxI\MM \GD^{-1}\MM\GD[\VE']\MM \ArgMax \notag\\
	& \neq & \GD^{-1}\MM\GD[\VE'] \MM \ArgMax \notag\Q,
\end{eqnarray}
then we know from \Thm{t1516X} that $\GD,\GD[\VE']$ and \ArgMax\ must together be unstable.
\Cf{Have we used any special properties of \ArgMax? \AStuff{Only that Argmax is deterministic.}}

But what have we gained?

The assumption $(\dagger)$ is if anything \emph{more} complex than the conclusion $\GD\MM P\,\ChanRef\,\GD[\VE']\MM P$ we used it to reach. And in both cases we still have to search through \GD's and \GD[\VE']'s to find an instance of $(\dagger)$.

\emph{What we've gained} is that showing an inequality $(\neq)$ is much easier --computationally-- than showing a refinement-failure $(\ChanRefN)$.
\Cf{In fact we need only search for only instances of
$\ArgMax\MM \ArgMaxI\MM\GD^{-1}\MM\GD[\VE'] \neq\GD^{-1}\MM\GD[\VE']$, since only then could $(\dagger)$ hold. Can we characterise $\GD^{-1}\MM\GD[\VE']$ for $\VE'{<}\VE$?
}